\documentclass[10pt,conference]{IEEEtran}
\IEEEoverridecommandlockouts
\usepackage{cite}
\usepackage{amsmath,amssymb,amsfonts}
\usepackage{graphicx}
\usepackage{textcomp}
\def\BibTeX{{\rm B\kern-.05em{\sc i\kern-.025em b}\kern-.08em
    T\kern-.1667em\lower.7ex\hbox{E}\kern-.125emX}}
    
\usepackage{hyperref}    
\usepackage[inline]{enumitem}
\usepackage{listings}

\usepackage{changepage}

\usepackage{subcaption}

\usepackage{ragged2e}
\usepackage{seqsplit}
\pdfmapfile{=rtxr.map} 

\usepackage[font=footnotesize,skip=1pt]{caption}
\DeclareCaptionFont{white}{\color{white}}
\DeclareCaptionFormat{listing}{%
  \parbox{\textwidth}{\colorbox{gray}{\parbox{\textwidth}{#1#2#3}}\vskip-4pt}}
\hypersetup{colorlinks,linkcolor={blue},citecolor={blue},urlcolor={blue}}

\usepackage{algorithm}
\usepackage{algpseudocode}
\makeatletter
\def\BState{\State\hskip-\ALG@thistlm}
\makeatother
\usepackage{balance}
\usepackage[skins,listings,breakable]{tcolorbox}

\makeatletter
\let\old@lstKV@SwitchCases\lstKV@SwitchCases
\def\lstKV@SwitchCases#1#2#3{}
\makeatother

\usepackage{lstlinebgrd}
\usepackage[utf8x]{inputenc}
\usepackage{ucs}
\usepackage{graphicx}
\usepackage{listings}
\usepackage{amsfonts}
\usepackage{amssymb}
\usepackage{tcolorbox}
\newtcbox{\highlight}[0]{boxsep=0pt,left=0pt,top=0pt,bottom=0pt,right=0pt,boxrule=0pt,arc=0pt,auto outer arc,colback=green,width=9cm}

\makeatletter
\let\lstKV@SwitchCases\old@lstKV@SwitchCases
\lst@Key{numbers}{none}{%
    \def\lst@PlaceNumber{\lst@linebgrd}%
    \lstKV@SwitchCases{#1}%
    {none:\\%
     left:\def\lst@PlaceNumber{\llap{\normalfont
                \lst@numberstyle{\thelstnumber}\kern\lst@numbersep}\lst@linebgrd}\\%
     right:\def\lst@PlaceNumber{\rlap{\normalfont
                \kern\linewidth \kern\lst@numbersep
                \lst@numberstyle{\thelstnumber}}\lst@linebgrd}%
    }{\PackageError{Listings}{Numbers #1 unknown}\@ehc}}
\makeatother

\lstdefinestyle{base}{
  language=Java,
  basicstyle=\ttfamily\scriptsize,
  frame=lines,
  keywordstyle=\color{blue}\textbf,
  commentstyle=\color[rgb]{0.0,0.4,0.0}\scriptsize,
  extendedchars=true,         
  breaklines=true   
  showspaces=false,
  showstringspaces=false, 
   numbers=left,
    stepnumber=1,   
        tabsize=1,
    breaklines=true,      
    xleftmargin={0.5cm},
  moredelim=**[is][\color{green}]{!!}{!!},
  moredelim=**[is][\color{orange}]{^}{^},
  moredelim=**[is][\color{red}]{@}{@},
   breakindent=0pt,                
     }  

\newcommand{\subparagraph}{}    

\newcommand*\circled[1]{\tikz[baseline=(char.base)]{
            \node[shape=circle,fill,inner sep=0.5pt] (char) {\textcolor{white}{#1}};}}
\usepackage{booktabs}

\usepackage[square,sort,comma,numbers]{natbib}

\usepackage{xcolor}
\usepackage{colortbl}
\usepackage{soul}

\usepackage{dblfloatfix}

\begin{document}
\title{What Distributed Systems Say: A Study of Seven Spark Application Logs\textsuperscript{\small *}
\\
\thanks{\textsuperscript{*}We give the title's credit to the influential Oliner and Stearley's paper, ``What Supercomputers Say: A Study of Five System Logs~\cite{oliner2007supercomputers}.''}
}

\author{\IEEEauthorblockN{Sina Gholamian}
\IEEEauthorblockA{\textit{University of Waterloo} \\
Waterloo, ON, Canada \\
sgholamian@uwaterloo.ca}
\and
\IEEEauthorblockN{ Paul A. S. Ward}
\IEEEauthorblockA{\textit{University of Waterloo} \\
Waterloo, ON, Canada\\
pasward@uwaterloo.ca}}

\maketitle

\begin{abstract}
Execution logs are a crucial medium as they record runtime information of software systems. 
Although extensive logs are helpful to provide valuable details to identify the root cause in postmortem analysis in case of a failure, this may also incur performance overhead and storage cost.
Therefore, in this research, we present the result of our experimental study on seven Spark benchmarks to illustrate the impact of different logging verbosity levels on the execution time and storage cost of distributed software systems. 
We also evaluate the log effectiveness and the information gain values, and study the changes in performance and the generated logs for each benchmark with various types of distributed system failures. 
Our research draws insightful findings for developers and practitioners on how to set up and utilize their distributed systems to benefit from the execution logs.    
\end{abstract}


\section{Introduction and Motivation}
The rapid growth of processing requirements and data scale in computing systems has contributed to the development and adaptation of large-scale, parallel, and distributed computation and storage platforms, \textit{e.g.}, Apache Spark and Hadoop Distributed File System (HDFS). 
Laterally, as the size of the data and computing systems grow, and they become more distributed in nature, evaluating their reliability and performance becomes more daunting.  
As such, execution log files and instrumentation of the source code are important origins of information for dependability analysis and gaining insight into the runtime state of the system. 
Execution logs have advantages over instrumentation, as they are readily available, do not require access to the source code, and do not introduce perturbation~\cite{malony1992performance}. 
However, instrumentation requires access to the source code, and it incurs perturbation due to the added instrumentation code.     

\begin{figure}[h]
\vspace{-3mm} 
\hspace*{-2mm}
\includegraphics[scale=.65]{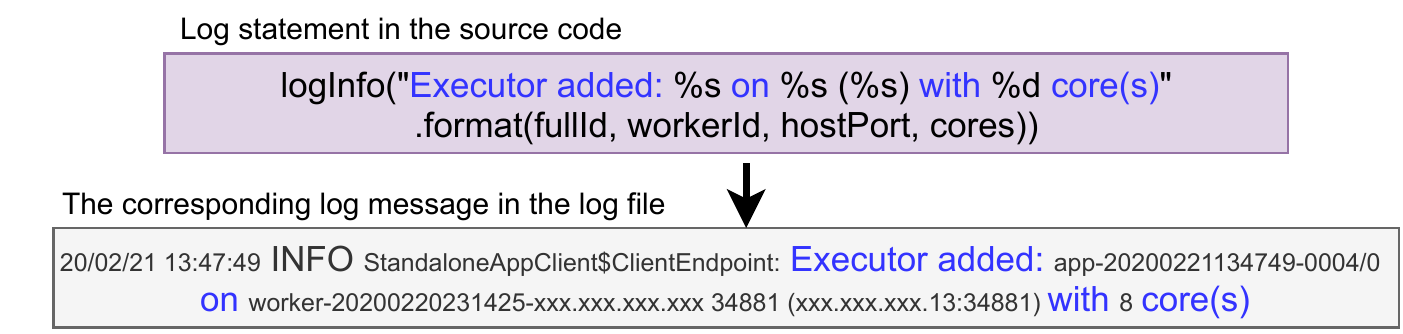}
\caption{Log statement and end product in the log file.\looseness=-1 }
\label{log_example}
\end{figure}

Logging is an important integral part of the software development process to record necessary run-time information~\cite{fu2014developers,hassan2008road}. 
Software developers insert logging statements into the source code to record a wealth of information such as variable values, state of the system, and error messages. 
Developers and system operators use this information for different purposes, among them failure and performance diagnosis~\cite{xu2009detecting,ding2015log2}. 
Although logging has proven benefits, it can incur system costs. 
Excessive logging can cause system overhead, such as CPU and I/O consumption. 
Contrarily, logging too little may miss important information and degrade the usefulness of execution logs~\cite{fu2014developers}. Authors of~\cite{he2018identifying} described a typical online system at Microsoft that could produce execution logs in the terabyte order-of-magnitude per day. 
As such, this high volume of logs can impair the quality of service for such systems. 
To address the trade-offs associated with the overhead of logging, well-known libraries, such as Apache Log4j~\cite{urllog4x} and SLF4j~\cite{urlself4j}, provide facilities for different levels and granularities of logging. 
The libraries provide different verbosity levels to dynamically control the number of logging statements being ultimately outputted to the log file on the storage medium. 
As each logging statement comes with a verbosity level, the logging library filters log messages by comparing the log statement's level with the dynamic log level specified by the user. 
Log4j has six verbosity levels available to the developers by default: \textit{fatal, error, warn, info, debug, and trace}. 
Figure~\ref{log_example} shows an example of a logging statement from Spark with \textit{info} verbosity level and its end product in the log file. 
In addition, each logging statement consists of a constant part, \textit{i.e.}, \textit{``Executor added: on with core(s)''}, and a variable part, \textit{i.e.}, \textit{``fullId''}.

Log levels represent a measure of the importance of the messages. 
For example, less verbose levels (\textit{i.e.}, \textit{fatal}, \textit{error}, and \textit{warn}) are used to warn the user when a potential problem happens in the system. 
On the other side, more verbose levels such as \textit{info}, \textit{debug}, and \textit{trace} are utilized to track more general system events and information or detailed debugging. 
Considering the flexibility that each log level brings, our goal in this research is to quantitatively measure the cost, in terms of storage, execution overhead, and information gain (IG) of log files while the distributed system is running under different log verbosity levels. 
Ultimately, we aim to reach a guideline on implications for developers and practitioners on how to utilize the logs in different verbosity level decisions while developing or operating distributed software systems in normal scenarios and in presence of failures. 
We guide our research with the following research questions (RQs):\looseness=-1 
\begin{itemize}
\item  \textbf{RQ1:}\textit{ what is the quantitative cost of logging in terms of computation time (CT) and storage overhead (SO)?} (\S\ref{rq1})
\item \textbf{RQ2:} \textit{how much information is gained from different log verbosity levels (VLs)?} (\S\ref{rq2})

\item \textbf{RQ3:} \textit{how the characteristics of logs change with distributed failures, \textit{i.e.}, distributed computation and storage failure? Does the entropy of logs increase when a failure happens?} (\S\ref{rq3})
\end{itemize}

For each RQ, we discuss the practical findings of our analysis and their implications for developers and practitioners on how to utilize the execution logs. 
Our research provides insight on how to choose the level of logging, and ultimately control the amount of generated logs and the information gain, and how the failures can be detected with entropy values. 
In addition, we provide a discussion on our findings and the implications for future improvements in distributed systems and their scheduling in case of system failures (\S\ref{discussion}).
With the motivation of helping developers and practitioners to gain more insight into the content of execution logs, and to make more deliberate logging level decisions, we pursue the following contributions in this paper: 
\begin{enumerate*}[label={\textbf{(\arabic*)}}]
\item We evaluate the performance and cost of logging for Spark under a set of batch and iterative workloads with different characteristics to calculate the overall execution time overhead and volume of generated logs (\textbf{RQ1}).
\item We calculate the information gained from the log files in different VLs based on their entropies and natural language processing (NLP) of logs with n-gram models and provide insights on how to make logging level decisions based on the observed cost and information gain from the log files (\textbf{RQ2}).
\item We introduce a comprehensive set of distributed system failures and evaluate the changes in execution log characteristics and entropy values, and provide insights on practical outcomes of our analysis for how to utilize execution logs to pinpoint failures (\textbf{RQ3}). Lastly, we release our labeled failure logs to encourage and enable further research in this field~\cite{datathispaper}.
\end{enumerate*}

\section{Approach and Setup}\label{sec_aproach}
\begin{figure}[h]
\vspace{-3mm}
\centering
\includegraphics[scale=.6]{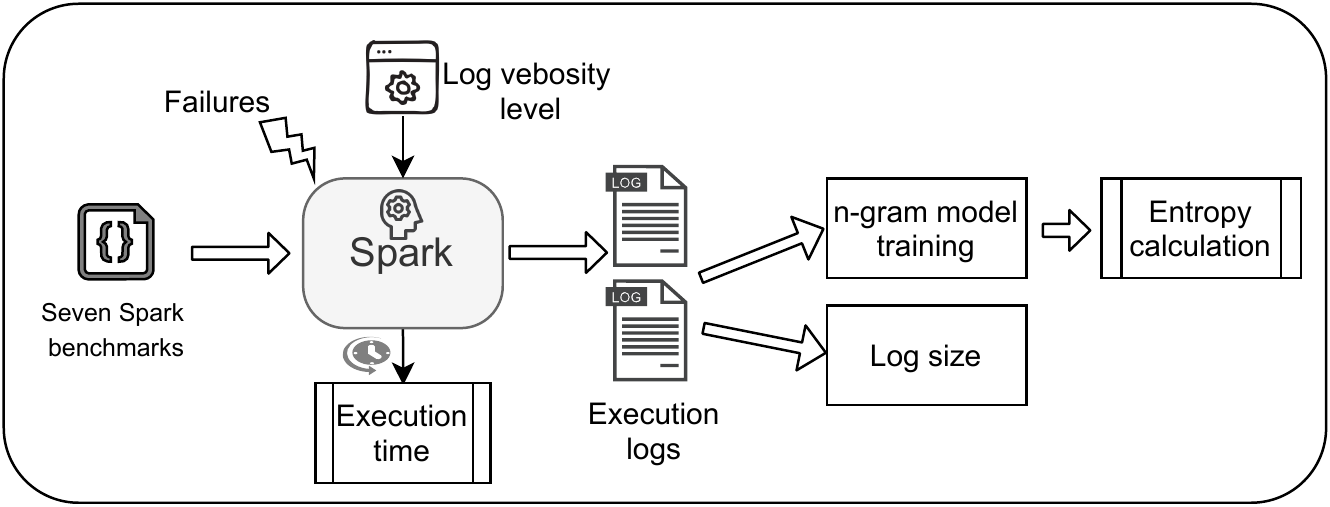}
\caption{Our approach for measuring the cost and effectiveness of the logs.\looseness=-1 }
\label{log_cost}
\vspace{-3mm}
\end{figure}

In this section, we present our approach and characterize the systems, their configurations, and the workloads that we use to conduct our study. 
Figure~\ref{log_cost} outlines the steps involved in our study. 
We categorize the logging cost into two system aspects: 1) execution overhead and 2) storage cost. 
We run seven Spark benchmarks with different log verbosity levels and calculate the execution times and the size of the generated logs. 
We then utilize Shannon's entropy theory~\cite{shannon2001mathematical} and n-gram models~\cite{ngrammodel} to measure the information gain by calculating entropies for different log levels with and without failures. 

\textbf{Apache Spark.} 
Since its introduction, Spark has been widely adopted as a big-data, distributed, and parallel processing framework. 
Spark builds upon Hadoop's MapReduce model and brings extra flexibility and improved performance. 
Additionally, Spark provides interfaces to other big-data platforms such as Hadoop's distributed file system, HDFS. 
To achieve higher performance and as an improvement to Hadoop MapReduce, Spark utilizes Resilient Distributed Datasets (RDDs)~\cite{zaharia2012resilient}, which retain the intermediate results in main memory, and therefore, reduces the overhead caused by the disk and network~\cite{zaharia2010spark}. 
This optimization benefits Spark the most in iterative tasks (\textit{e.g.}, Transitive Closure), as the following stages of the task rely on the intermediate results from the prior stages. 
Because of its improved performance and widespread use, we deploy a Spark cluster to perform our study.\looseness=-1

\begin{figure}[h]
\vspace{-4mm}
\centering
\includegraphics[scale=.6]{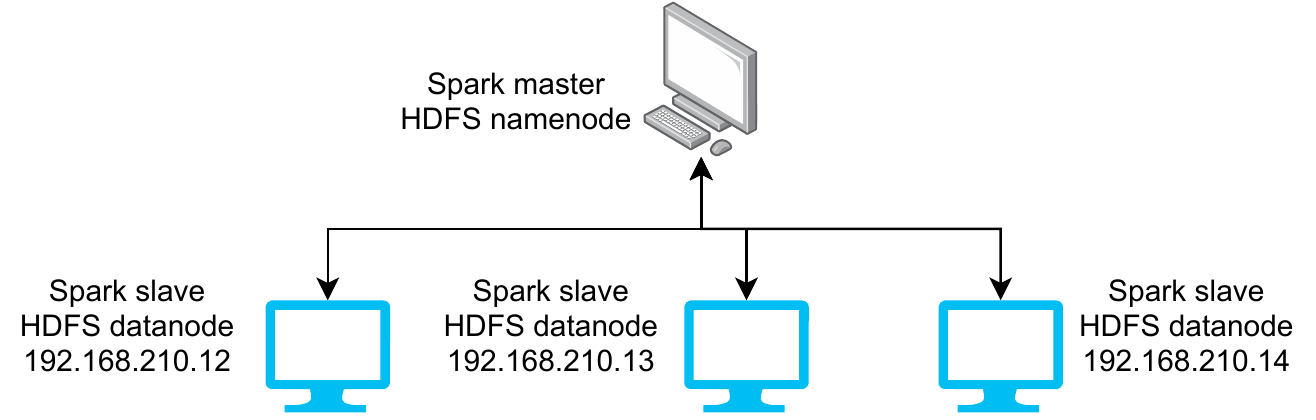}
\caption{Design of the distributed cluster, consisting of one master/name node and three slave/data nodes.}
\label{cluster_setup}
\vspace{-2mm}
\end{figure}

\textbf{Hardware.}
Table~\ref{cluster_tabel} and Fig.~\ref{cluster_setup} show the main hardware characteristics and the architecture of our deployed cluster, respectively. 
Each node in the cluster has 12 (12*2 hyper-threaded) cores, 32 GBs of memory, and 2 storage disks of 1 TB each.
We evaluate the benchmarks on a cluster of 4 commodity machines illustrated in Fig~\ref{cluster_setup}. 
Each machine is equipped with dual 2.40GHz Intel Xeon E5-2620 CPUs, supporting a total of 24 hyperthreads per machine and a 1Gbps NIC. 
All servers run Ubuntu Server 16.04.6 LTS 64-bit with kernel version 4.4.0-159-generic. 
{\renewcommand{\arraystretch}{1.1}
\begin{table}[h!]
\scriptsize
    \vspace{-1mm}
  \begin{center}
    \begin{tabular}{p{.6cm}|p{1.8cm}|p{1.1cm}|p{.8cm}|l|p{1cm}}
      \textbf{Name} & \textbf{Role} & \textbf{Cores} & \textbf{Memory} & \textbf{Disk} & \textbf{Local IP}\\ 
      \toprule
     styx01 & Master/NameNode  & 12 (24 HT)  & 64 GB & 2*1 TB&{\tiny192.168.210.11}\\ 
 
    styx02 &  Worker/DataNode  & 12 (24 HT)  & 64 GB & 2*1 TB&{\tiny192.168.210.12} \\ 
    styx03 &  Worker/DataNode  & 12 (24 HT)  & 64 GB & 2*1 TB & {\tiny192.168.210.13}\\ 
    styx04 &  Worker/DataNode  & 12 (24 HT)  & 64 GB & 2*1 TB & {\tiny192.168.210.14}\\ 
      
    \end{tabular}
  \end{center}
  \vspace{-1mm}
   \caption{Styx cluster for Spark computation and HDFS.}
    \label{cluster_tabel}
    \vspace{-2mm}
\end{table}
}

\textbf{Framework setup.} 
In our experiments, we use Spark 2.4.4 and Hadoop 2.9.2. 
We use \textit{styx01} as a dedicated server for the Spark \textit{master} and HDFS \textit{name node}, while having one Spark \textit{slave} and HDFS \textit{data node} on each of the three other machines, \textit{styx02, styx03, and styx04}. 
Hadoop file system block size is 128 MBs and replication is set to 3. 
We set up each Spark slave to use 12 available cores and up to half of the available memory \textit{(i.e.}, 32 GBs out of 64 GBs) on each of the nodes. 
The frameworks have been carefully configured according to their corresponding user guides and the characteristics of the system (\textit{e.g.}, number of CPU cores and memory size). 
Table~\ref{spark_settings} summarizes the related configurations for HDFS and Spark nodes.\looseness=-1
\begin{table}[t]
    \label{spark_settings}

\centering
\begin{tabular}{|c||c|c|}
\hline
    Framework & Parameter &Value \\ \hline
    HDFS & block size& 128 MBs \\ 
    HDFS & Replication factor& 3 \\
    Spark & Worker/Executor per node &1\\ 
    Spark & Cores per executor & 24 HT\\
    Spark & Memory per executor& 32 GBs\\
    Spark & Driver memory& 10 GBs\\
     \hline
\end{tabular}  
   \caption{Main Spark and HDFS settings.}
    \label{spark_settings}
    \vspace*{-5mm}
\end{table}

\textbf{Benchmarks.}
In this research, we experiment on seven different Spark benchmarks and provide a brief explanation of each one in the following: \begin{enumerate*}[label=\protect\circled{\arabic*}]
\item \textbf{WordCount (WC)}, which counts the number of times each word appears in the input dataset. 
By applying transformations such as \textit{`.reduceByKey()'} on RDDs, WordCount outputs a dataset of (word, value) pairs, saved to a file on HDFS. 
WordCount is a popular Spark's benchmark that allows us to assess CPU and I/O costs associated with different levels of logging. 
\item \textbf{TeraSort (TS),} which sorts randomly generated rows of key-value (KV) pairs with each KV being 100 bytes. 
The TeraSort implementation and its random data generator engine, TeraGen, are both adopted from~\cite{urlterasort}.

\item \textbf{TransitiveClosure (TC),} checks and implements linear transitive closure (LTC) on a graph, iteratively.  
For example, if \textit{x, y, and z} are three vertices, and \textit{(x,y)} and \textit{(y,z)} represent edges between \textit{x and y}, and \textit{y and z}, respectively, for satisfying the transitive closure property, a new edge is added between \textit{x and z}.  
LTC grows paths by one edge, by joining the graph's edges with the already-discovered paths in each iteration. 
TC is an iterative CPU-intensive workload. 

\item \textbf{PageRank (PR)} is an iterative graph algorithm that ranks URLs by considering the number and rank of URLs referring to it. 
For example, the more URLs with higher ranks refer to a URL under consideration (URLUC), the URLUC's rank increases. 
For the PageRank's implementation, we use the implementation provided with the Spark's example package.
\item \textbf{TestDFSIO (DF),} which is a benchmark designed to evaluate the I/O (read/write) performance by using Spark's tasks to read and write multiple files in parallel. 
The benchmark aims to read and write an even amount of data to HDFS on each node in the cluster.  The implementation is adapted from~\cite{urltestdfsio}.

\item \textbf{GradientBoostingClassificationTrees (GC),} which is a machine learning algorithm for classification, that generates a prediction model as an ensemble of decision trees. 
In this use case, the number of classes is set to two, the depth of the trees is set to five, and we perform 200 iterations for the model training. 

\item \textbf{LinearDiscriminantAnalysisClustering (LD),} which implements LDA clustering algorithm, \textit{i.e.}, unlabeled data, that clusters the input documents into three different topics.
\end{enumerate*}

Table~\ref{table_benchmark} summarizes the benchmarks used in the experiments, along with their characterization such as CPU or I/O (disk and network) intensive, and if the computation happens iteratively. 
The sizes of the input datasets are also shown, and we refer to the benchmarks with their abbreviation in the rest of the paper, as shown in Table~\ref{table_benchmark}.
During the benchmark selection, we were deliberate to include a variety of workloads such as Spark's conventional benchmarks, \textit{e.g.}, WC and PR, and machine learning ones, \textit{e.g.}, GC and LD.

\begin{table*}[t]
\footnotesize
\centering
\begin{tabular}{p{3.5cm} | p{3.8cm}| p{2cm}|p{6.5cm}} 
 \toprule
 \rowcolor{blue!15}\textbf{Benchmark (abbrv.)} & \textbf{Task type} & \textbf{Input data size}& \textbf{Notes} \\ [0.5ex] 
 \midrule
 WordCount (WC) & CPU and I/O intensive  & 52 GBs & The output is pairs of (word, count) written to HDFS. \\
 \midrule 
  TeraSort (TS) & Iterative, I/O, and CPU intensive & 2 GBs & Sorts randomly generated (key, value) pairs, and the size of each pair is 100 bytes.\\ [1ex] 
  \midrule 
  TransitiveClosure (TC) & Iterative and CPU intensive& small (few KBs) & Calculates the transitive closure on a randomly generated graph with 200 edges and 100 vertices. \\ [1ex] 
    \midrule
   PageRank (PR) & Iterative and CPU intensive & 40 MBs & Ranks web pages based on their popularity.\\ [1ex] 
     \midrule 
    DFSIO (DF) & I/O intensive& 20 $\times$ 1 GB &  Writes and then reads 20 files of one GB each to HDFS. \\ [1ex]
      \midrule 
   GradientBoostingClassification Trees (GC) & Iterative and CPU intensive & small (205 KBs)&  Trains 200 decision trees with the depth of five for classification of a decision problem, \textit{i.e.}, yes (1) or no (0). \\ [1ex] 
     \midrule
   LinearDiscriminantAnalysis Clustering (LD) & Iterative and CPU intensive & 21 MBs & Clusters the input data into three topics using LDA. \\ [1ex] 
  \bottomrule
\end{tabular}
\caption{Benchmark characteristics.}
\label{table_benchmark}
\vspace{-2mm}
\end{table*}

\begin{figure*}
\begin{subfigure}{0.135\textwidth}
   \includegraphics[width=\linewidth]{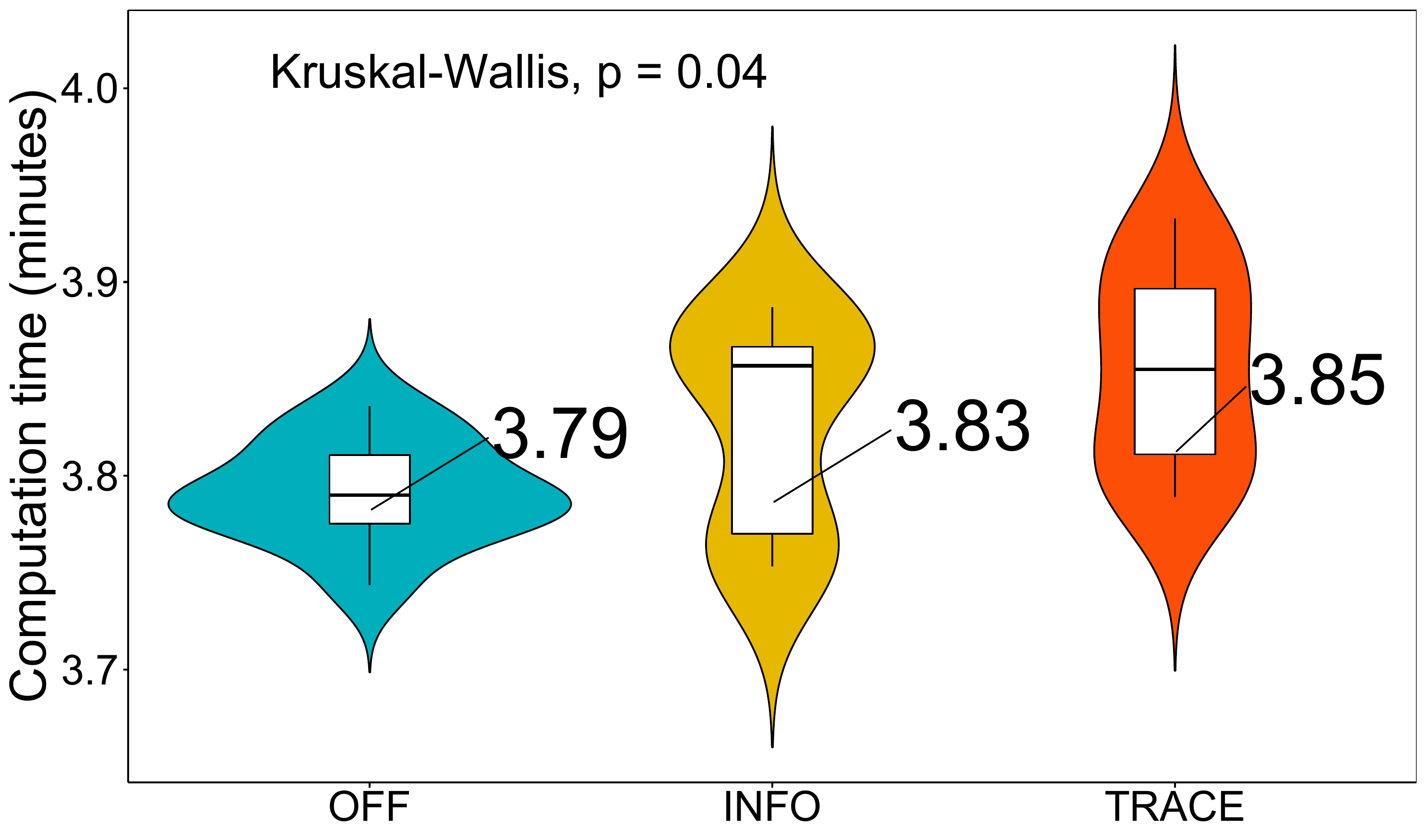}\vspace*{-2mm}
   \caption{WC CT.} \label{wc_ct}
\end{subfigure}
\begin{subfigure}{0.138\textwidth}
   \includegraphics[width=\linewidth]{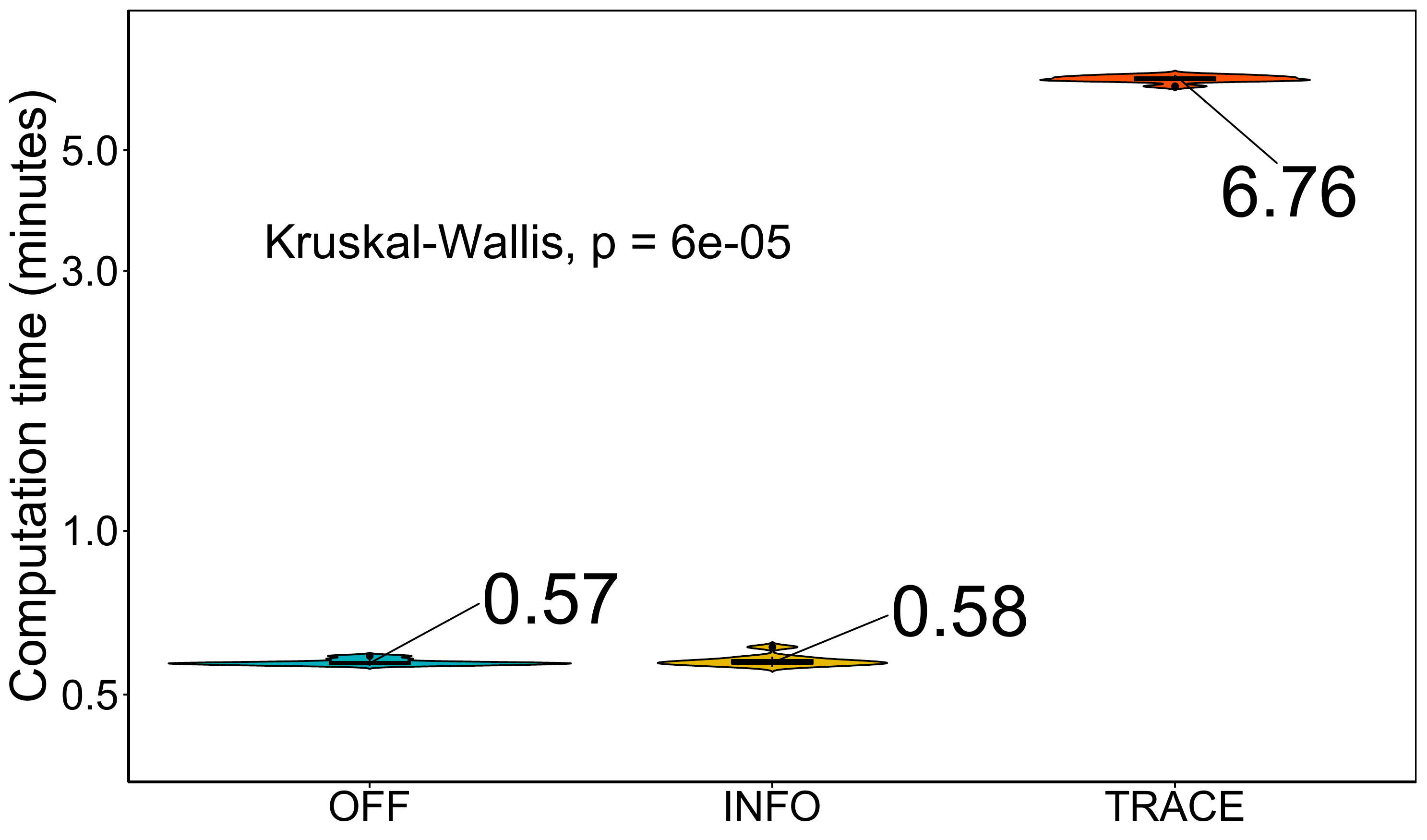}\vspace*{-2mm}
   \caption{TS CT.} \label{ts_ct}
\end{subfigure}
\begin{subfigure}{0.138\textwidth}
   \includegraphics[width=\linewidth]{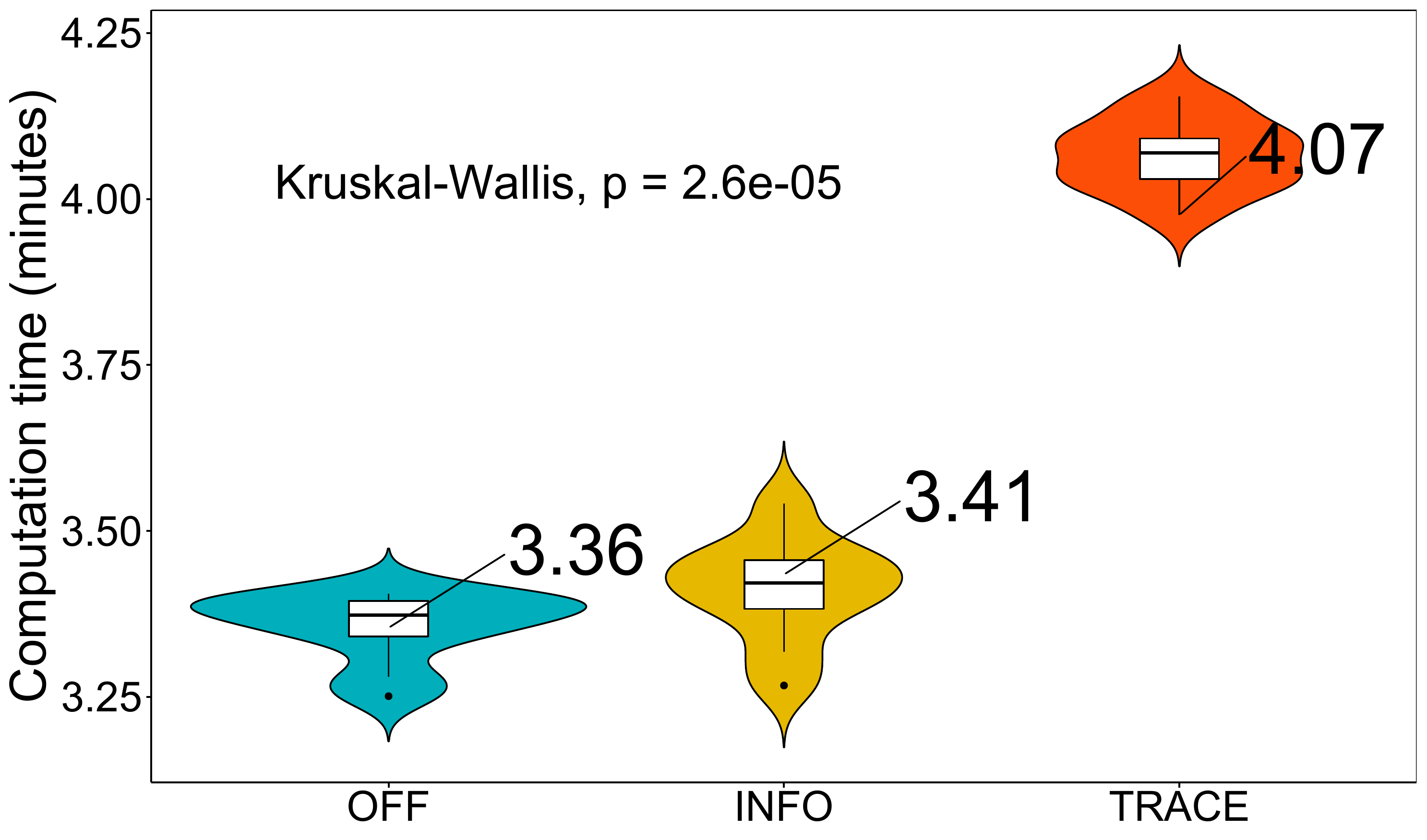}\vspace*{-2mm}
   \caption{TC CT.} \label{tc_ct}
\end{subfigure}\hspace*{\fill}
\begin{subfigure}{0.138\textwidth}
   \includegraphics[width=\linewidth]{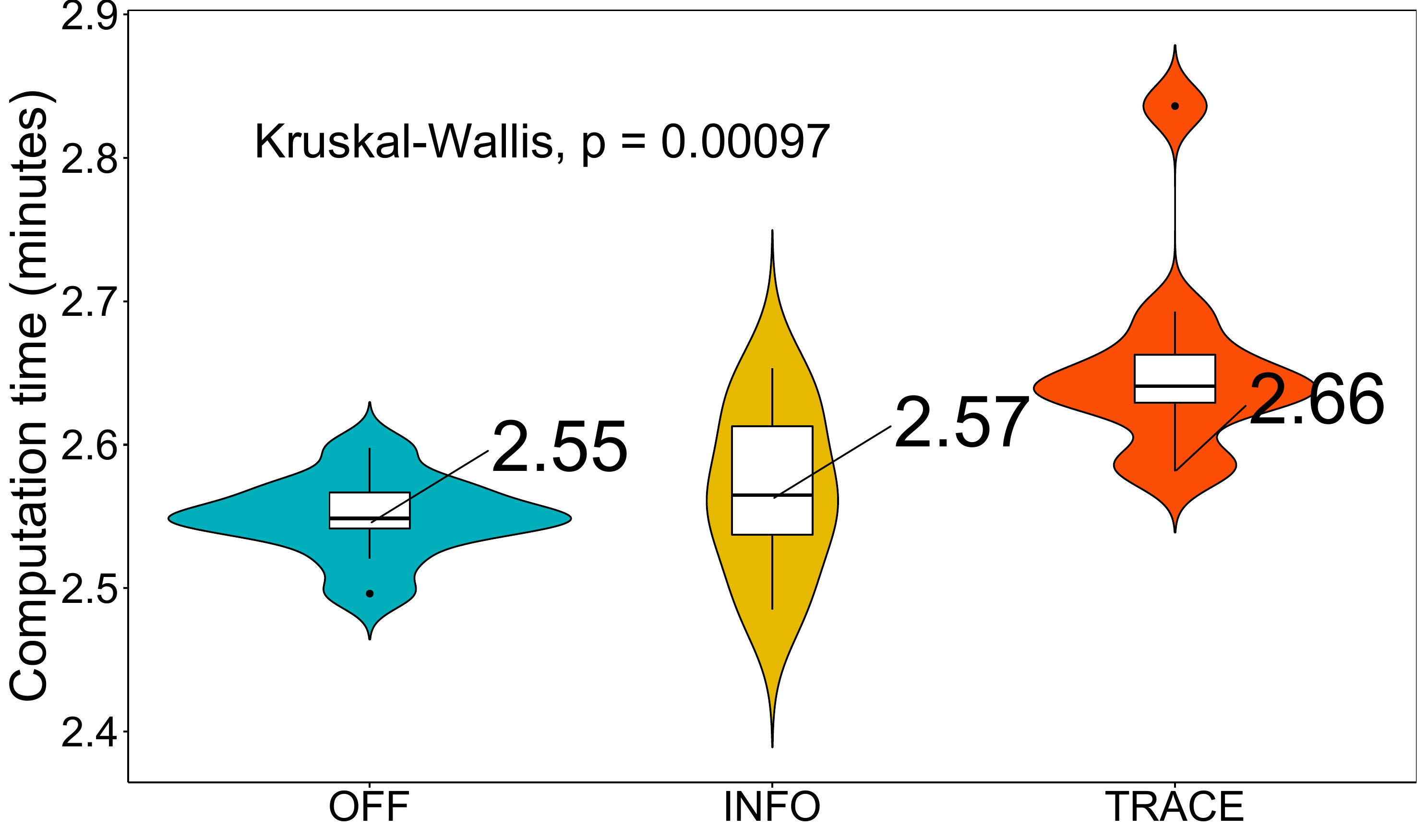}\vspace*{-2mm}
   \caption{PR CT.} \label{pr_ct}
\end{subfigure}
\begin{subfigure}{0.138\textwidth}
   \includegraphics[width=\linewidth]{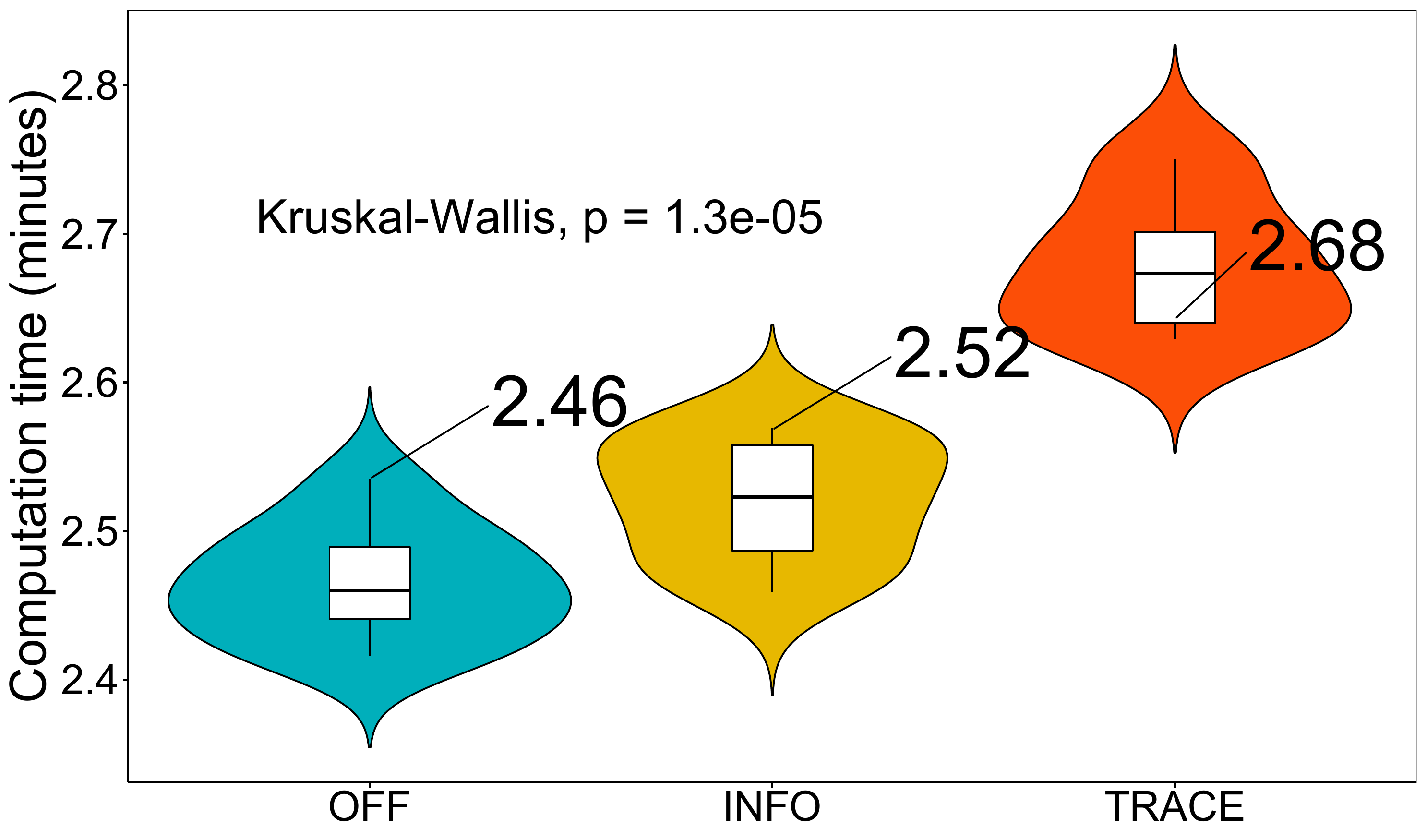}\vspace*{-2mm}
   \caption{DF CT.} \label{df_ct}
\end{subfigure}
\begin{subfigure}{0.138\textwidth}
   \includegraphics[width=\linewidth]{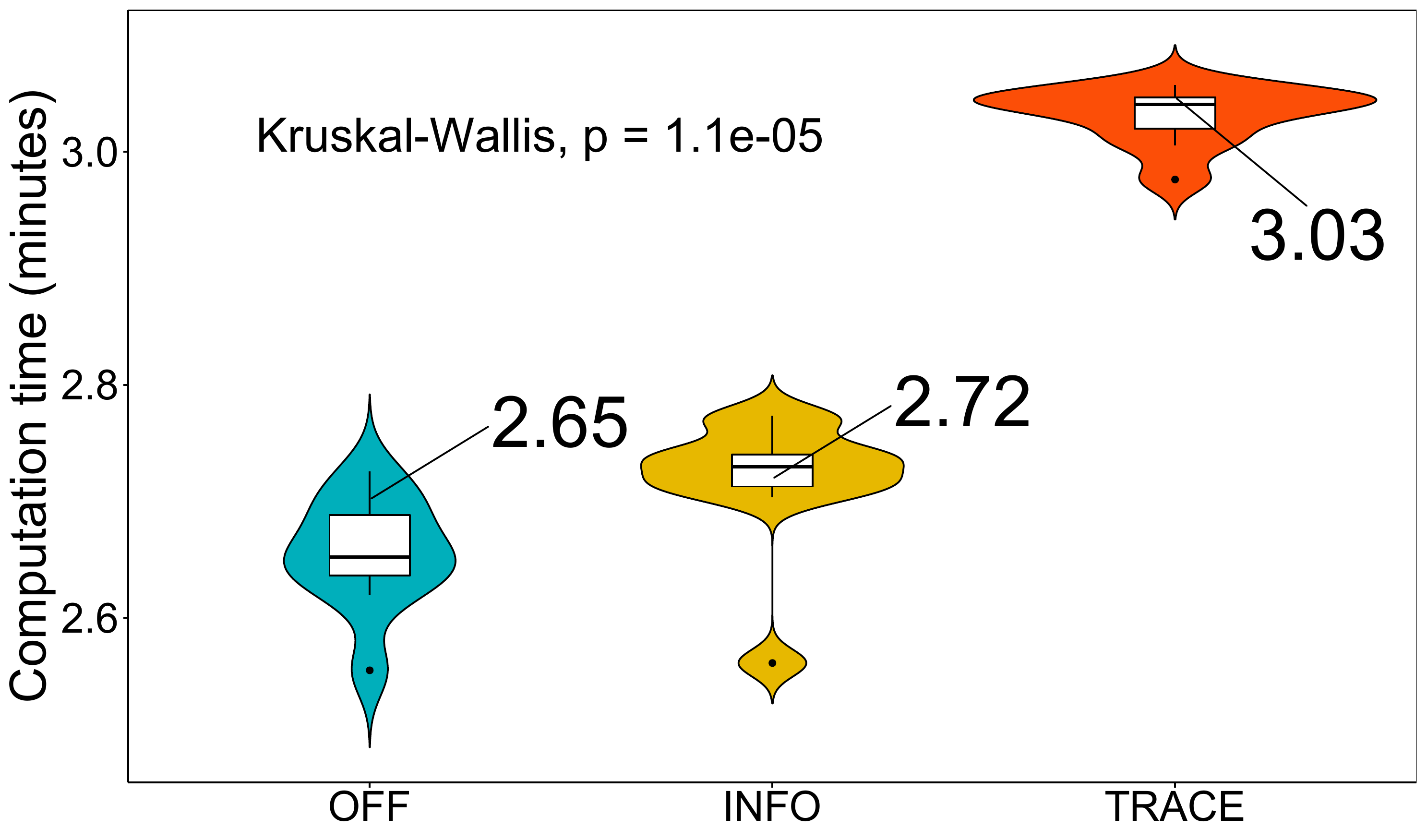}\vspace*{-2mm}
   \caption{GC CT.} \label{gc_ct}
\end{subfigure}
\begin{subfigure}{0.138\textwidth}
   \includegraphics[width=\linewidth]{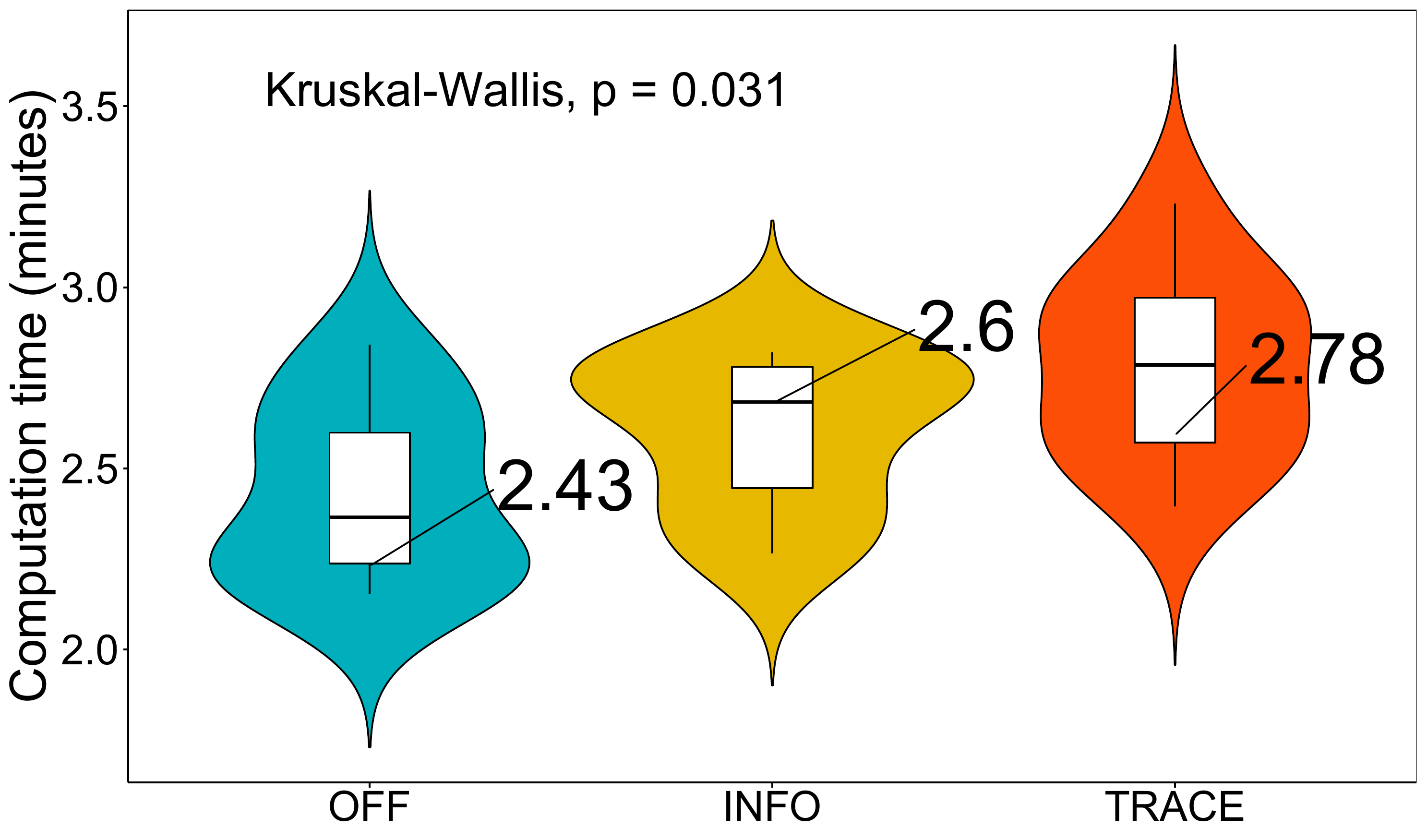}\vspace*{-2mm}
   \caption{LD CT.} \label{ld_ct}
\end{subfigure}

\begin{subfigure}{0.14\textwidth}
   \includegraphics[width=1\linewidth]{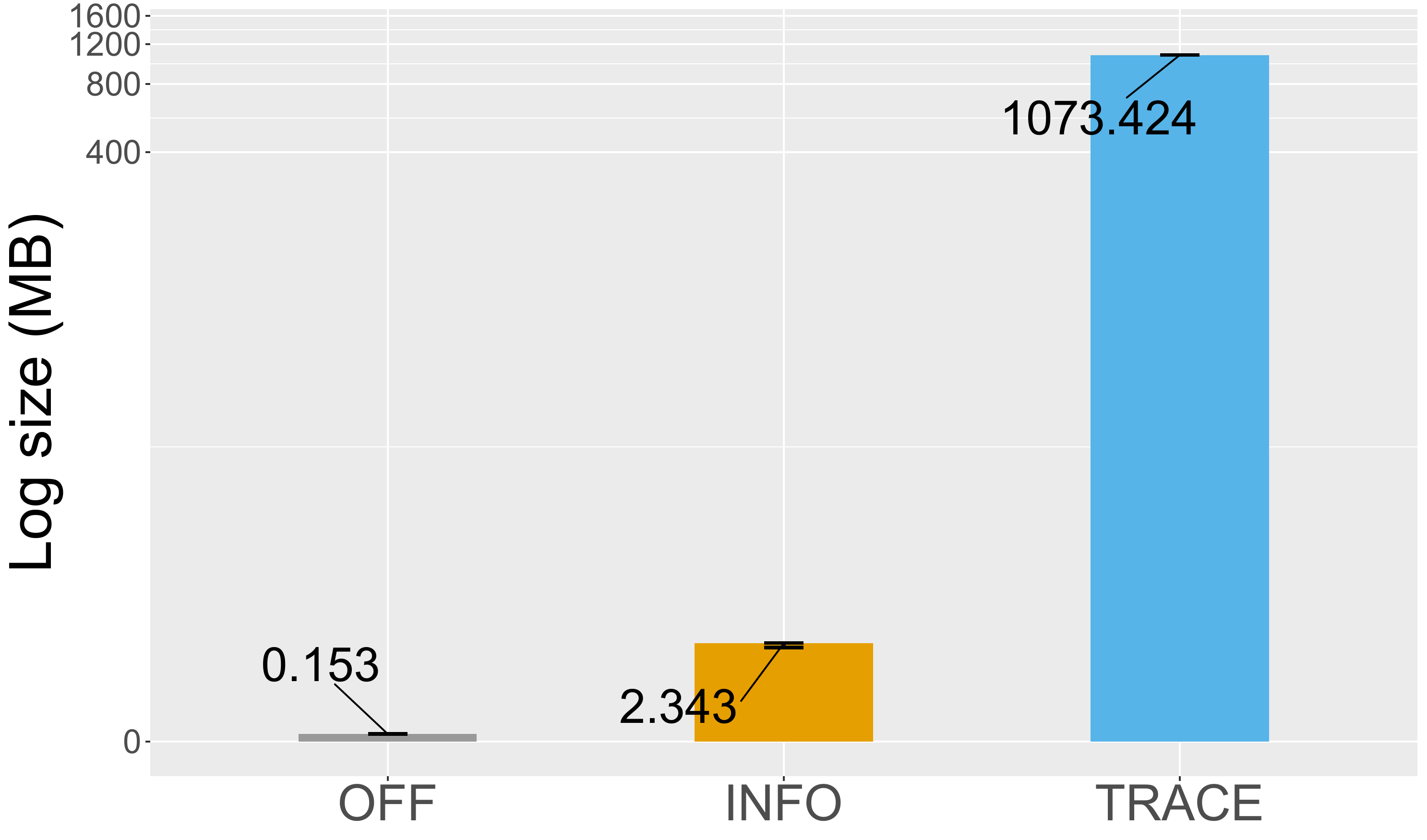}\vspace*{-1mm}
   \caption{WC SO.} \label{wc_ls}
\end{subfigure}\hspace*{\fill}
\begin{subfigure}{0.14\textwidth}
   \includegraphics[width=1\linewidth]{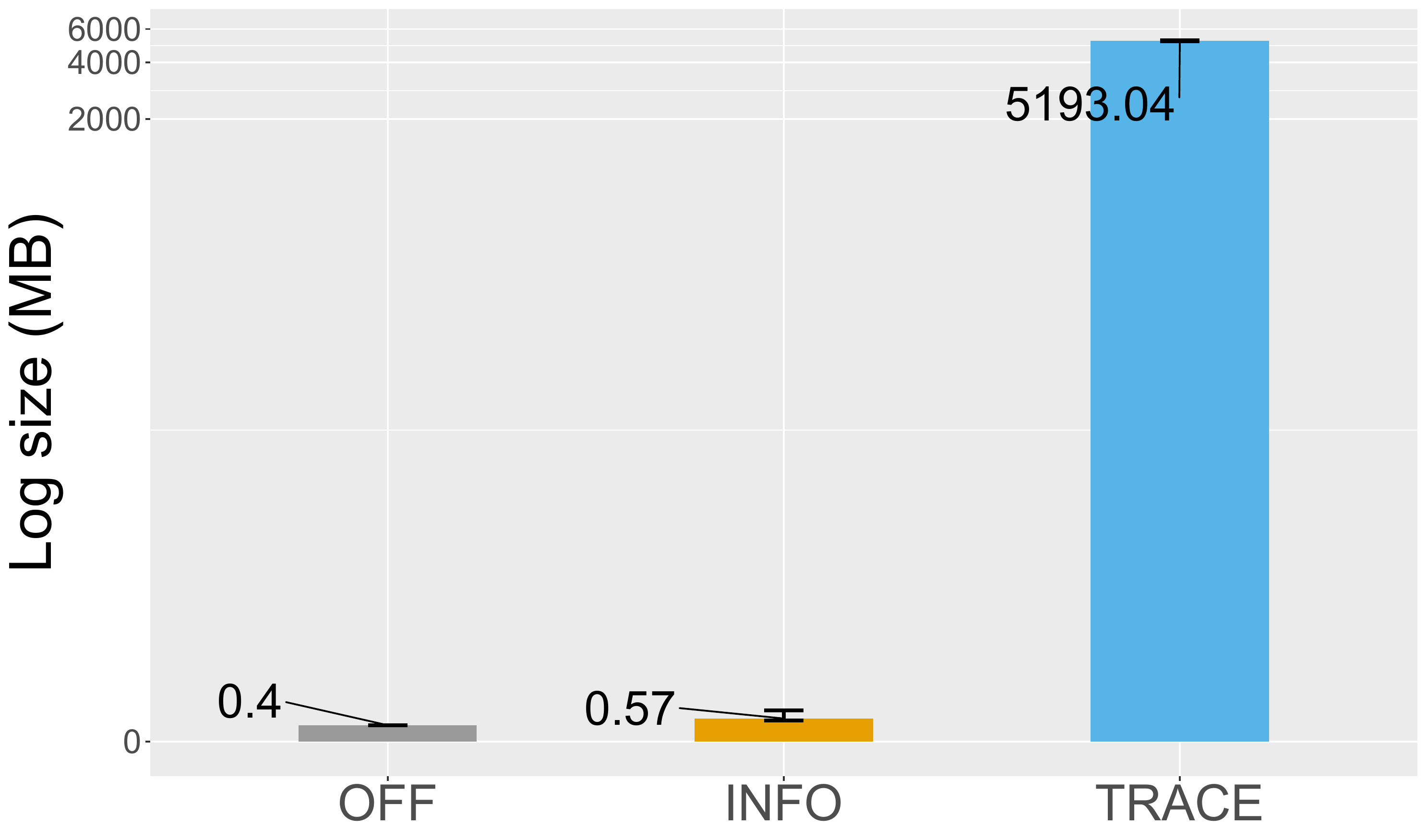}\vspace*{-1mm}
   \caption{TS SO.} \label{ts_ls}
\end{subfigure}\hspace*{\fill}
\begin{subfigure}{0.14\textwidth}
   \includegraphics[width=1\linewidth]{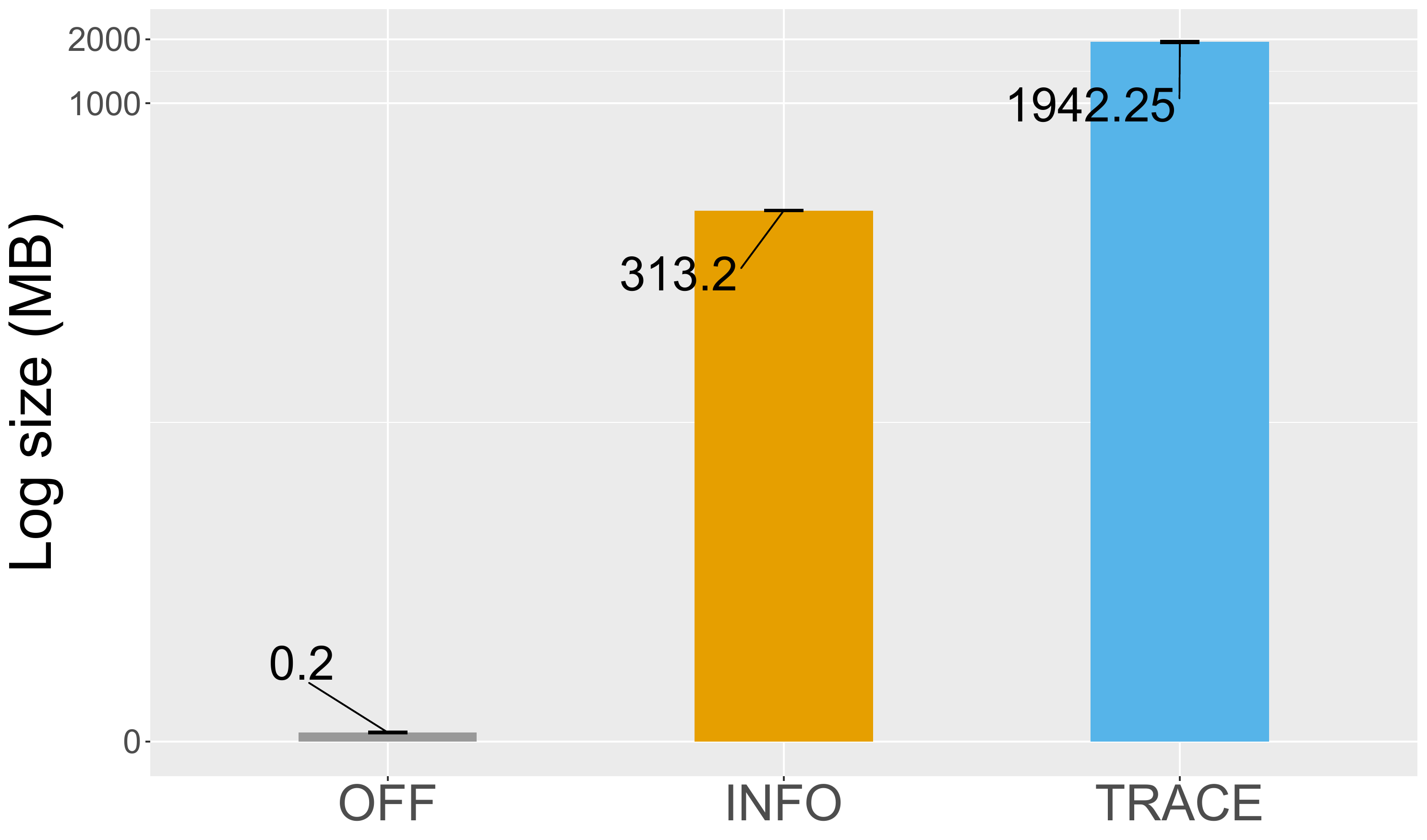}\vspace*{-1mm}
   \caption{TC SO.} \label{tc_ls}
\end{subfigure}\hspace*{\fill}
\begin{subfigure}{0.14\textwidth}
   \includegraphics[width=1\linewidth]{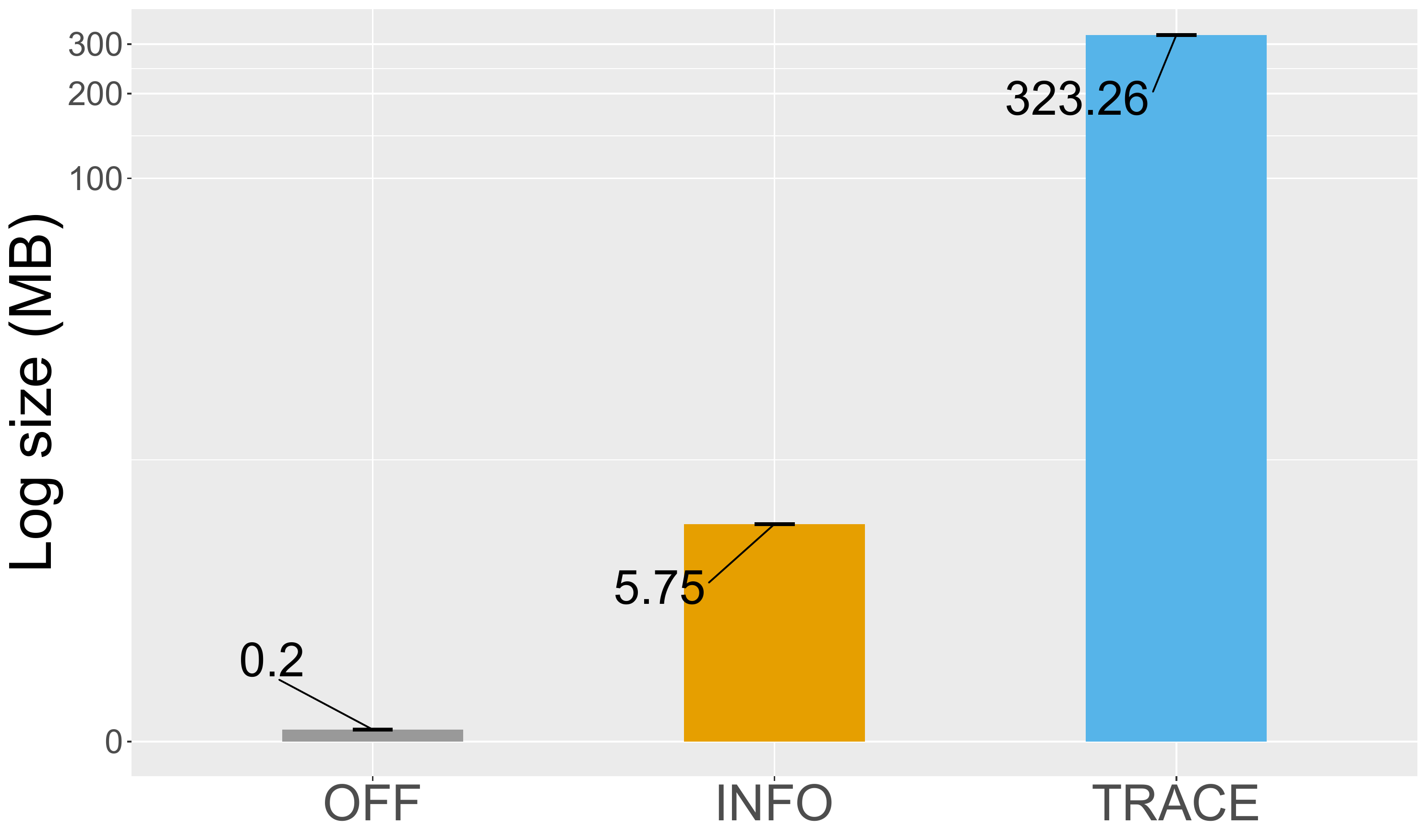}\vspace*{-1mm}
   \caption{PR SO.} \label{pr_ls}
\end{subfigure}\hspace*{\fill}
\begin{subfigure}{0.14\textwidth}
   \includegraphics[width=1\linewidth]{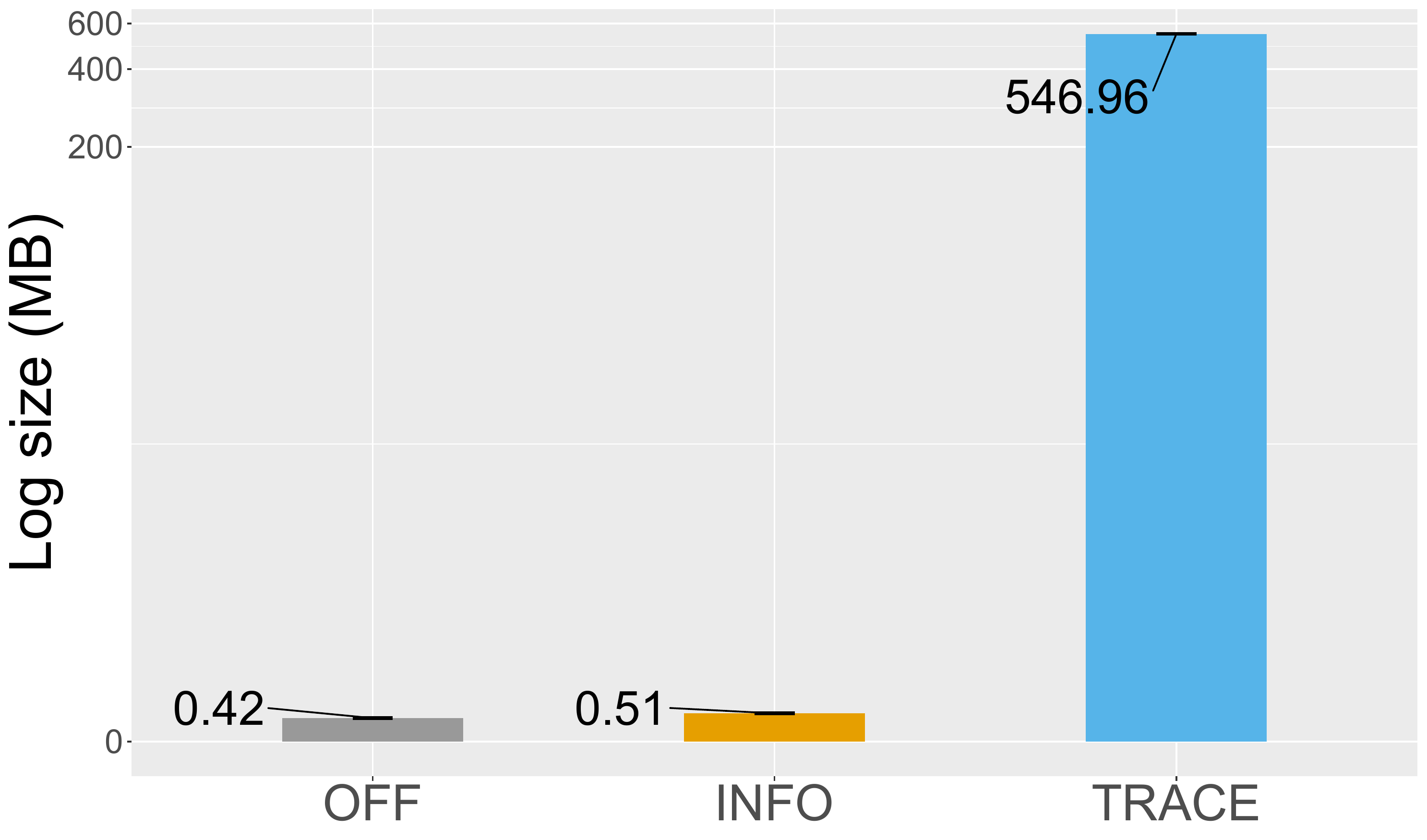}\vspace*{-1mm}
   \caption{DF SO.} \label{df_ls}
\end{subfigure}\hspace*{\fill}
\begin{subfigure}{0.14\textwidth}
   \includegraphics[width=1\linewidth]{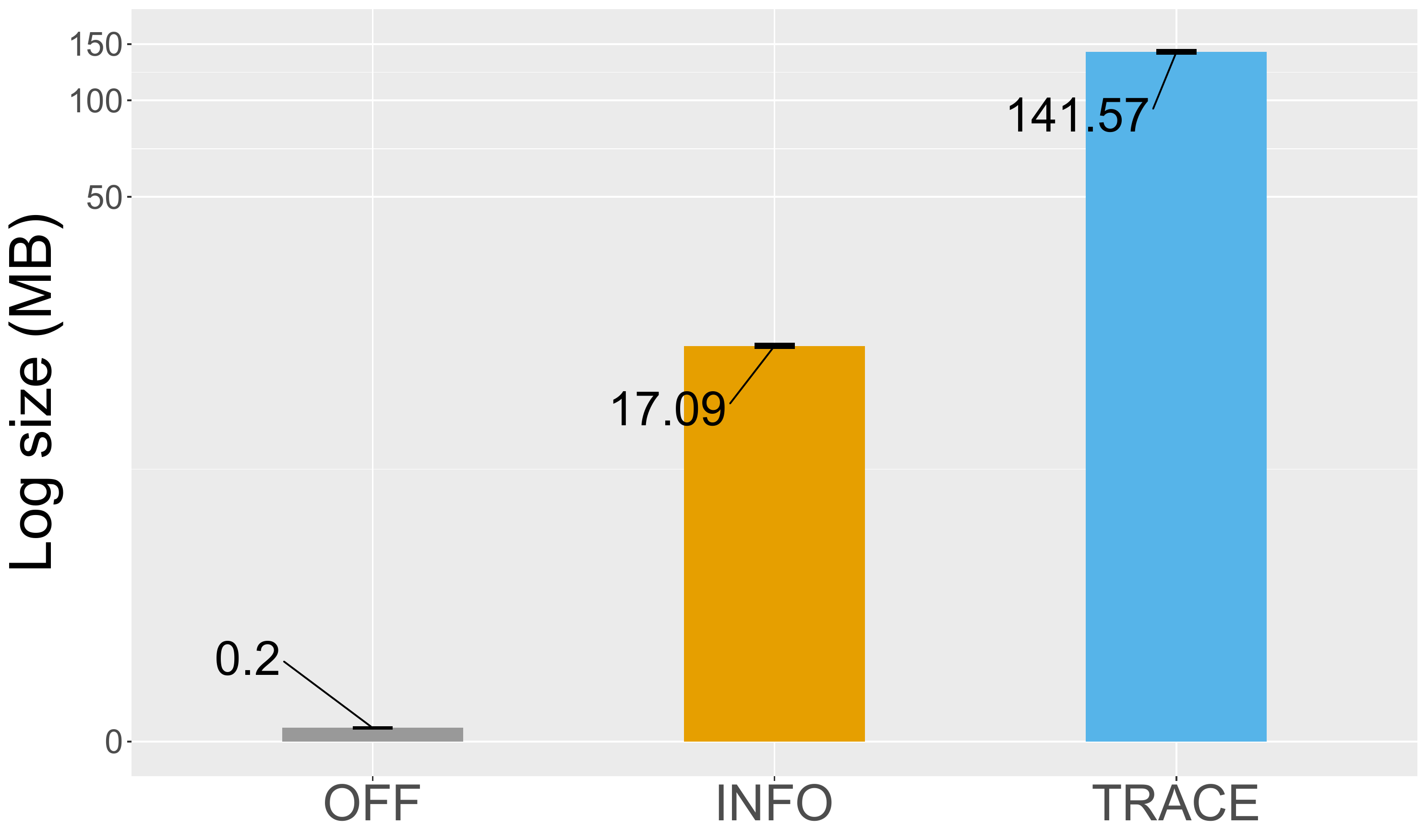}\vspace*{-1mm}
   \caption{GC SO.} \label{gc_ls}
\end{subfigure}\hspace*{\fill}
\begin{subfigure}{0.14\textwidth}
   \includegraphics[width=1\linewidth]{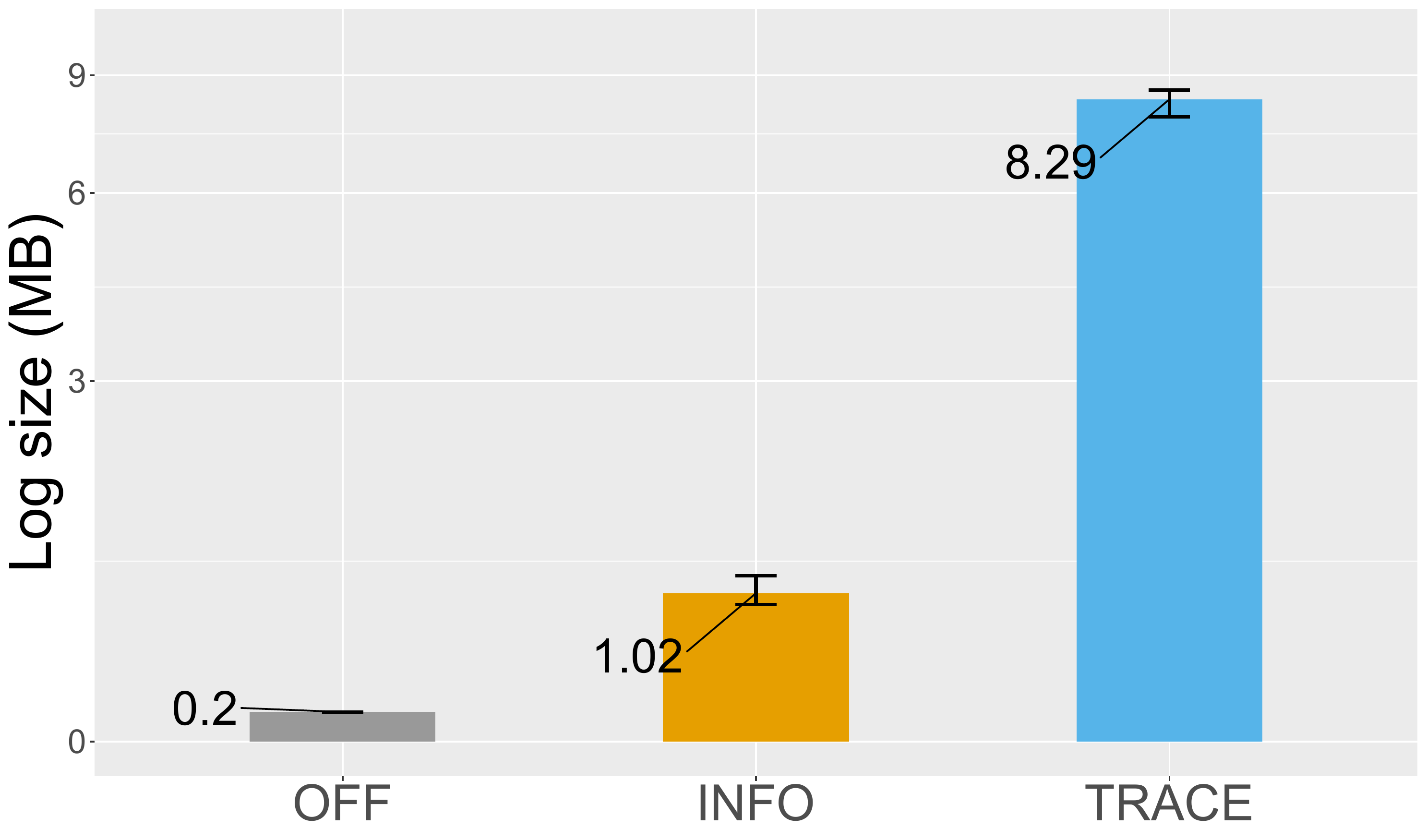}\vspace*{-1mm}
   \caption{LD SO.} \label{ld_ls}
\end{subfigure}\hspace*{\fill}
\vspace{1mm}
\caption{Computation time (CT) and storage overhead (SO) for different benchmarks. }\label{ct_vs_so}
\vspace{-2mm}
\end{figure*}

\section{RQ1: Cost of Logging}\label{rq1}
There exists various qualitative and quantitative metrics for assessing logging cost. 
Quantitative metrics consider the overhead of logging on different subsystems of the computing systems, \textit{e.g.}, CPU and I/O overhead~\cite{ding2015log2}, and qualitative metrics assess the cost of logging in terms of developer and user experience, such as the cost of revealing private information through logs~\cite{li2020qualitative}.
In our work, we focus on quantitative measurement of the logging cost and for this purpose, we conduct a set of experiments to evaluate the impact of logging verbosity level on the size of the generated logs, as well as the effect on the performance of the Spark. 
To measure the cost of logging, we run multiple Spark benchmarks on our commodity cluster and calculate the logging cost in terms of the size of the generated log and the execution time for each benchmark. 
The measured computation time and storage values in this section serve as a baseline for comparison on further RQs.

\subsection{\textbf{Computation time (CT)}} 
Figures~\ref{wc_ct}-\ref{ld_ct} show the violin chart and interquartile range with mean values (noted in text) for execution time of different benchmarks. 
We also perform Kruskal-Wallis test~\cite{kruskal1952use} to reject the null hypothesis and ensure statistically significant values, \textit{i.e., $p\leq 0.05$}.
Violin chart improves on boxplot chart by providing the width in the graph as the density of points in the experiments.
The vertical axes represent the execution time in minutes, and the horizontal axes show different verbosity levels (VLs). 
We select to show \textit{info} and \textit{trace} VLs as the former is generally enabled by default and presents the normal mode of logging, and the latter represents the maximum amount of logging which is widely utilized during failure diagnosis~\cite{chen2004failure,zhou2019latent}. 
This provides a picture of the lower and upper bounds of logging. 
Besides, we present the data for logging \textit{off} to present the other end of the spectrum (least amount of logging) compared to the \textit{trace} level (most amount of logging). 
The overall trend shows the more verbose VLs result in higher execution times. 
Because \textit{trace} VL enables more detailed logging and executes additional lines of code (and more I/O system calls), it incurs a minor but noticeable execution time overhead across different benchmarks, when compared to \textit{info} level.

\subsection{\textbf{Storage overhead (SO)}}
Figures~\ref{wc_ls}-\ref{ld_ls} show the size of generated execution logs for different benchmarks. 
The vertical axes represent the size of the log file in MB, and the horizontal axes show different verbosity levels. 
The sizes show the aggregated logs for all the nodes (\textit{i.e.,} master and workers) of the Spark cluster. 
Similar to computation times, \textit{trace} level enables more detailed logging, which results in significantly higher volumes of log data when compared with less verbose log levels. 

\textbf{Results review.} Based on the conducted experiments, we observe that the execution time of the benchmarks increases as the log level becomes more verbose.
This observation is congruent with the knowledge that the execution of logging messages infers extra CPU, I/O, and storage cost. 
Overall, excluding TS, we see on average 8.01\% overhead in execution time when the \textit{trace} log level is enabled versus the \textit{info} level. 
For the log file size, we notice on average a $\sim$268X increase in the volume of the generated logs for the \textit{trace} log level versus the \textit{info} verbosity level. 
We did a further investigation to better understand the $\sim$12X increase in computation time for TS, and we observed the significant amount of generated logs for TS in \textit{trace} level when compared to \textit{info} (\textit{i.e.}, 5 GBs \textit{vs.} 0.5 MBs). 
Because TS is a CPU-intensive benchmark, we rationalize that its CT suffers noticeably due to the significant amount of logs outputted in the \textit{trace} level. 
Comparing CT and SO values for different benchmarks, we observe that the amount of generated logs in different VLs is benchmark dependent, and CPU-intensive applications (\textit{e.g.}, TS, TC) observe a higher slowdown due to more verbose logging. 
 
\textbf{SO mitigation.} Prior work has shown that due to the high level of repetitiveness in log files, they can benefit from large compression ratios, up to 84\%~\cite{yao2020study}. 
Therefore, the noticeable difference in storage cost can be mitigated by the compression of log files to $\sim$43X.

\begin{tcolorbox}[breakable, enhanced]
\small \textbf{Finding} \textbf{1.} \textit{Overall, we observe on average  8.01\% and  $\sim$268X overhead in the execution time and storage when the \textit{trace} log level is enabled versus the \textit{info} level, respectively, and CPU-intensive workloads suffer more from a higher degree of logging. }

\textbf{Implications.} \textit{Considering the trade-offs, if the worst-case 8.01\% execution time is acceptable, by utilizing log rolling, compression, and continuous achieving, the storage overhead of more verbose logging can be further lowered.}
\end{tcolorbox}

\subsection{RAM Disk}
We used the hard disk drive (HDD - TOSHIBA MG04ACA200E - 7,200 RPM) as the storage medium for collecting the logs. 
Because we observed significant degradation for the performance of some of the CPU-intensive applications such as TeraSort when the \textit{trace} level is enabled, we further investigate the impact of utilizing faster storage systems for log collection. 
Because developers and practitioners mostly utilize trace log level for debugging, we rationalize that the debugging data can be saved in memory temporarily to expedite the debugging process and the final debugging outcome can be transferred to the disk when the debugging is finalized. 
In addition, although memory storage is volatile and there is a risk of debugging data loss due to power outage, we presume this risk is manageable as we are concerned with debugging data in contrast to the actual execution logs, and the experiments can be repeated in case the debugging data is lost. 
Additionally, as new storage technologies become faster, \textit{e.g.,} solid-state drive (SSD), and its latency edges closer to the main memory speed, this data point shows the maximum potential improvement that comes in from a faster storage paradigm, \textit{i.e.}, a latency lower bound and a \textit{`hypothetical'} storage medium that is as fast as the main memory. 
Table~\ref{ramdisk_ct} compares the CT values for HDD versus \textit{RAM Disk} for benchmarks that we observed a noticeable increase in CT when \textit{trace} level is enabled. 
RAM Disk is a utility that allows us to map a portion of RAM as disk space and redirect benchmark logs to the space on RAM.
Our goal is to show how much of the extra CT introduced because of the slow storage medium can be recovered by leveraging a faster storage medium, assuming non-volatile storage mediums become as fast as RAMs. 

{\renewcommand{\arraystretch}{1.2}
\begin{table}[h]
 \centering
\begin{tabular}{|c||c|c|c|c|c|c|}
\hline
    CT (min) &        TS    & TC    & PR & DF   & GC   &     LD    \\ \hline
    \textbf{HDD}&     7.40  & 4.07  &  2.66  & 2.68 & 3.03 & 2.78       \\ \hline 
    \textbf{RAM Disk}& 5.73 & 3.99  &  2.61  & 2.64 & 2.99 & 2.68  \\
    \hline
     \textbf{CT reduction (\%)}& 22.62 & 2.05 &  2.02   & 1.38  &  1.19 & 3.60 \\
     \hline
\end{tabular}  

   \caption{Computation time values for RAM Disk \textit{vs.} HDD  for \textit{trace} level.}
    \label{ramdisk_ct}
\end{table}
}

\begin{tcolorbox}[breakable, enhanced]
\small\textbf{Finding 2.}\textit{ TeraSort, which generates a significantly higher amount of logs in trace level compared to info level, shows the highest CT reduction while using RAM Disk.}

\textbf{Implications.} \textit{Faster storage mediums can mitigate some of the overhead associated with logging for CPU-intensive workloads that generate a significant amount of logs in more verbose log levels.}
\end{tcolorbox}

\section{RQ2: Log Effectiveness}\label{rq2}
In RQ2, we evaluate the relationship between the log verbosity levels and their effectiveness. 
Although more verbose logs are used generally for debugging, there has not been any effort to quantitatively assess the effectiveness of logs in more verbose levels.  
In other words, although the common perception is that a higher degree of logging translates to more effectiveness of the logs, this assumption might not completely hold true.   
For this purpose, we introduce a new metric for calculating the effectiveness of logs based on entropy values and investigate whether or not more verbose VLs are more effective. 

\textbf{Log effectiveness (LE).} 
LE is a quantitative measure of logs' effectiveness in achieving their goals, which is mainly problem diagnosis and troubleshooting. 
For example, Yuan \textit{et al.}~\cite{yuan2012characterizing} showed that in their experiments when log statements exist, developers could diagnose system problems \textit{2.2X} faster compared to not having the logs. 
In this study, LE is directly related to Entropy (\textit{i.e.}, $LE \propto Entropy$) that we clarify in the following.
As illustrated in Figure~\ref{log_example}, log statements consist of two parts: static and dynamic content. 
Static content of log statements originates from the source code, and dynamic content is the value of variables that are printed in the log files as the system is running. 
As such, the dynamic content of the logs can be different in each iteration, whereas the static part is unchanged and has the same value in every iteration of the program. 
Therefore, for more verbose VLs to be more effective than less verbose ones, they should result in higher Entropy values and information gain (IG), as this signifies more unique runtime content. 
In other words, higher dynamic content translates to more runtime information and value of variables, which is positively related to higher values of entropy, IG, and LE:

\begin{flushleft}
\centering
\noindent\fcolorbox{black}{blue!2}{
\textit{\small  ($\uparrow$Dynamic content) $\to$ ($\uparrow$Entropy) $\to$ ($\uparrow$IG) $\to$ ($\uparrow$LE) }
}
\end{flushleft}

\textbf{Shannon's entropy}. 
We use entropy as a metric to measure the dynamic content and effectiveness of the log records. 
Shannon’s entropy~\cite{shannon2001mathematical} is used to measure the amount of information that is contained in an information source (\textit{e.g.}, a text file). 
Entropy is calculated as: \colorbox{blue!10}{$H=-\sum_{n=1}^{N} p(i)\log_2 p(i)$}, where $p(i)$ is the probability of a possible character happening in the log data~\cite{shannon2001mathematical}. 
The more random (\textit{i.e.}, less repetitive) the content of the log file, the higher the entropy. 
For the purpose of experimentation, we focus on the \textit{info} and \textit{trace} log levels as they are used during the deployment and development of the software, respectively, to gain insight from the end user and developer perspectives. 
Because Spark generates only a few MBs of logs in \textit{info} level for some benchmarks and to perform the experimentation on equal log sizes, we randomly sample 1 MB size of Spark's logs for both \textit{trace} and \textit{info} levels for each benchmark and measure the entropies.
Table~\ref{log_effectiveness} shows the entropy values per character for log files in \textit{trace} and \textit{info} verbosity levels. 
The character-level entropy values are slightly higher for the \textit{trace} log level, which partially signifies higher information gain (IG) and less repetitiveness for this level.\looseness=-1 
{\renewcommand{\arraystretch}{1.2}
\begin{table}[h]
 \centering
\begin{tabular}{|c||c|c|c|c|c|c|c|}
\hline
    Entropy &     WC & TS&TC &PR &DF& GC&LD  \\ \hline
    \textbf{Info} & 5.24& 5.31 & 5.16  &5.21& 5.26  & 5.23 &5.26\\ \hline 
    \textbf{Trace} & 5.41&5.39  &5.38 &5.40&  5.37 & 5.34  &5.40\\
     \hline
\end{tabular}  
   \caption{Shannon's entropies for \textit{info} and \textit{trace} for various applications.}
    \label{log_effectiveness}
\end{table}
}

\textbf{N-gram model.} Although character level entropy explains the randomness of single characters in logs, it does not provide insight on the sentence level repetitiveness of log messages. 
It is more reasonable to calculate the entropies for a sequence of words, as log statements are inserted as a sequence of tokens (\textit{i.e.}, words and variables) in the source code.   
To accommodate for a sequence of words, which bears higher semantic meanings for log messages, entropy is also used for a sequence of grams (\textit{i.e.}, words or tokens), such as calculating the probabilities of a sequence of tokens in the English language. 
For this purpose, prior research has suggested the use of n-gram models~\cite{hindle2012naturalness}, to capture the repetitiveness of a sequence of words. 
The n-gram model captures the probability distribution of the log data, and once trained, it can predict the probability distribution of the next token in new log sequences by utilizing order-n Markov model approximation. 
This approximation considers the probability of $i_{th}$ element in the sequence of $n$ tokens to be predicted based on $n-1$ preceding tokens~\cite{jurafsky2000speech}. 
Therefore, we can estimate the probability of $a_i$ succeeding tokens $a_{i-1}, a_{i-2}, ..., a_{i-n+1}\vspace{1mm}$ with: 
\begin{equation}\small 
p(a_i|a_{i-1}a_{i-2}...a_{i-n+1}) = \frac{count(a_{i}a_{i-1}a_{i-2}...a_{i-n+1})}{count(a_{i-1}a_{i-2}...a_{i-n+1})} 
\end{equation}
Based on this model, the entropy for a sequence of tokens is: 
\begin{equation}\label{char_entropy}
H=- \frac{1}{N}\sum_{n=1}^{N} log \hspace{1mm} p(a_i|a_{i-1}a_{i-2}...a_{i-n+1})
\end{equation}

\begin{figure}
\centering
\includegraphics[width=.99\linewidth]{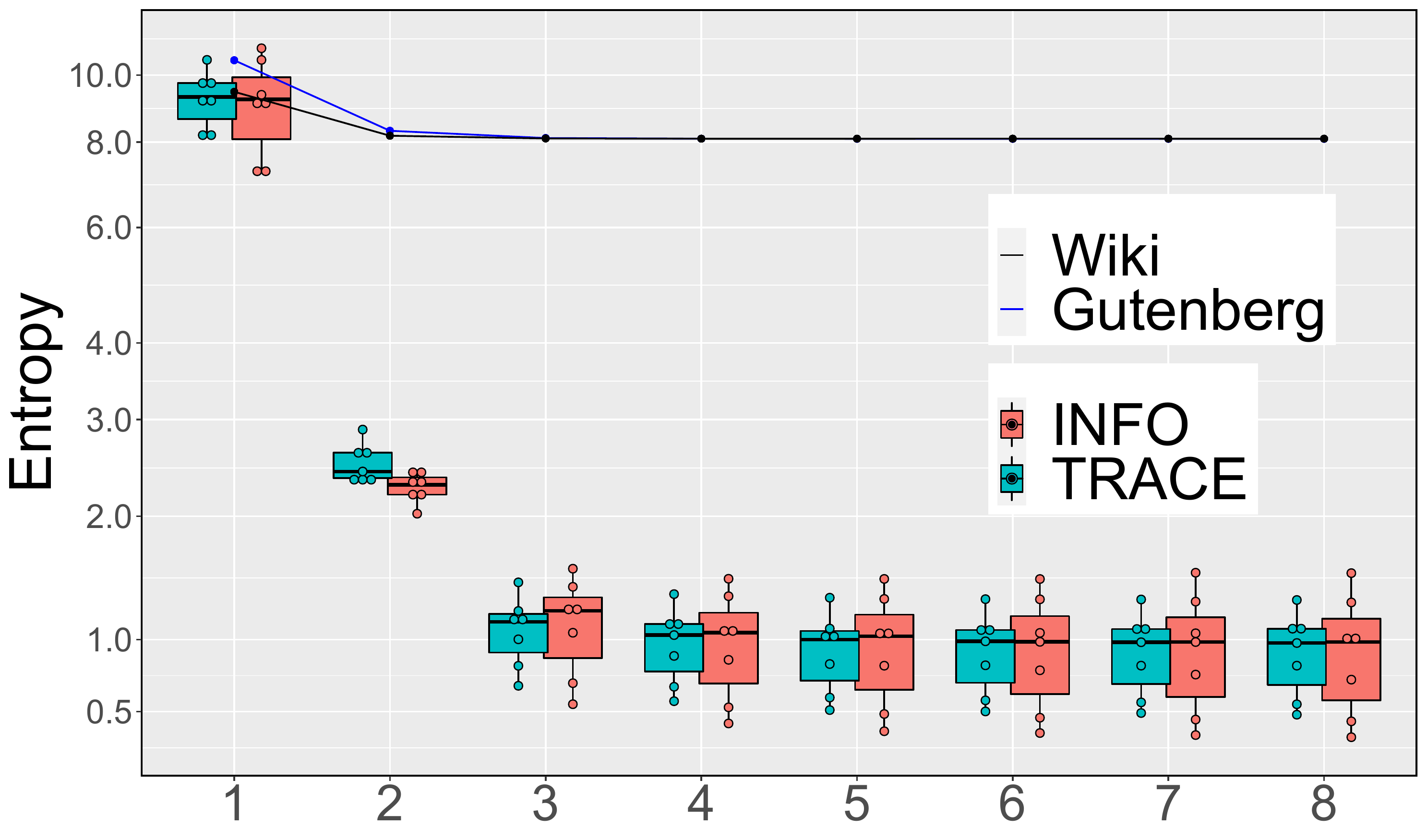}
\caption{Entropy for n-gram models for Spark logs and English text.}
\label{entropy_nlp}
\vspace{-3mm}
\end{figure}

To measure the sentence-level information gain from the logs, we evaluate the entropy of a sequence of log tokens in both \textit{info} and \textit{trace} levels with n-gram models and compare it with common English text such as \textit{Gutenberg}~\cite{urlgutenburg} and \textit{Wiki}. 
Gutenberg is a collection of English books, and Wiki is the English articles from Wikipedia. 
To train and test the n-gram models on the sequences of logs, we randomly sample 1 MB of data from each benchmark and perform a 90\%-10\% train-test split. 
We run ten-fold cross-validation to avoid overfitting~\cite{ng1997preventing} and plot the average entropies of 10 iterations for n-gram models in the range of $n \in (1,8)$ in Figure~\ref{entropy_nlp}. 
English text entropies stabilize around eight as the size of n-gram increases, and the median values for \textit{trace} and \textit{info} stabilize at 0.975 and 0.982, respectively. 
Our experiment reveals that English text has higher entropy than log files, and hence it has less repetitiveness, which is also observed in prior research on the naturalness of software artifacts~\cite{hindle2012naturalness,tu2014localness,gabrilovich2009wikipedia,gholamian2021naturalness}. 
Lower baseline entropy of software logs compared to natural language text is beneficial as it results in \textit{`distinguishable'} entropy changes while detecting anomalous log lines (\textit{i.e.}, peaks in entropy values)~\cite{gholamian2021naturalness}, which can be utilized for log failure detection. 
Interestingly enough, our comparison shows, in Spark's case, the n-gram entropies are comparable for \textit{trace} level when compared to the \textit{info} level.    
This suggests that although \textit{trace} level logging results in a higher volume of logs, \textit{trace} log sequences are \underline{\textit{not necessarily}} less repetitive, and \textit{trace} \underline{does not} benefit from noticeable {\textit{higher information gain}}, as IG from an event (\textit{e.g.}, log event) is directly related to its entropy~\cite{pal1991entropy}. 
In addition, a higher amount of repetition that results in larger log files might decrease their effectiveness. 
Redundancy is an undesirable feature of logs since it adds noise to the log files and complicates the understanding of the program's behavior and hinders problem diagnosis through logs~\cite{yuan2012characterizing,hassan2008industrial}.

\begin{tcolorbox}[breakable, enhanced]
\small \textbf{Finding 3.} \textit{Although trace level generates a larger volume of logs, trace data does not provide a noticeable higher entropy, and hence, does not necessarily carry higher information gain and effectiveness when compared to less verbose log levels.} 

\textbf{Implications.} \textit{We presume \textit{trace} logs show comparable IG to \textit{info} because they contain higher repetition rather than unique dynamic values.}
\end{tcolorbox}

\section{RQ3: Failure Assessment}\label{rq3}
Logs are widely utilized in failure detection and performance diagnosis~\cite{xu2009detecting,ding2015log2}. 
Therefore, in this section, we study the effectiveness of the information gain approach in system failure detection. 
To evaluate the effect of system failures on the generated logs, we design a framework to inject different types of distributed failures and measure their impact on logs and how the IG approach can be applied to extract log lines related to the failures. 
As numerous failure scenarios exist, our goal is not to provide a comprehensive list, but to investigate common failures in a distributed environment.  
We categorize the distributed failures in four main categories:
\begin{enumerate}
\item \textbf{Compute node failure} happens when a compute resource becomes unavailable. 
We synthesize this scenario by terminating one of the Spark's worker nodes.  
\item \textbf{Storage node failure} in a distributed environment happens when a storage medium becomes unavailable. 
As we utilize HDFS with the replication factor of three, the integrity of the data remains intact in case of a single node failure, however, the latency of reads and writes to the storage will increase for some compute nodes that require access to data on non-local HDFS nodes. 
We synthesize this failure by terminating one of the HDFS data nodes.   
\item \textbf{Communication interference}, which resembles a scenario in the distributed network with variable latency and a probability of packet loss. 
This category can be initially observed as a performance degradation, and eventually may lead to a complete failure if the communication delay between distributed compute and storage nodes surpasses a system's predefined timeout.  

\item \textbf{Combined failure} resembles a scenario in which multiple nodes become unavailable simultaneously for various reasons such as power outages. 
We simulate this scenario by terminating a cluster node that hosts both Spark compute and HDFS storage nodes.    
\end{enumerate}  

With this failure categorization, our goal is to observe the changes in the content of the logs files and apply information gain approaches to detect failures. 
\ul{The hypothesis is that we should observe a higher information gain during a failure, as the failure related logs should resemble different dynamic content.}   
As such, we evaluate the entropy of logs during their normal and abnormal (\textit{i.e.}, failure) time intervals. 
Because we noticed comparable entropy values for \textit{info} and \textit{trace} log levels (Figure~\ref{entropy_nlp}), in the following, we focus on \textit{info} verbosity level as it is the default log level during the deployment, and, additionally, storage overhead of logs becomes more manageable. 
We run each Spark's benchmark in \textit{info} level for ten iterations with and without the aforementioned failures and evaluate the changes in execution time and the storage overhead for the generated logs. 
In addition, we also evaluate the changes in information gain (entropy) with the normal and failure logs.  
For entropy calculation, the n-gram model is trained on the normal execution runs (\textit{i.e.}, without failures) and tested on runs with failures. We choose $n=5$ for the n-gram model, as according to Figure~\ref{entropy_nlp}, entropy values are stabilized for $n\geq 5$.\looseness=-1

\begin{figure*}
\centering
\includegraphics[width=.99\linewidth]{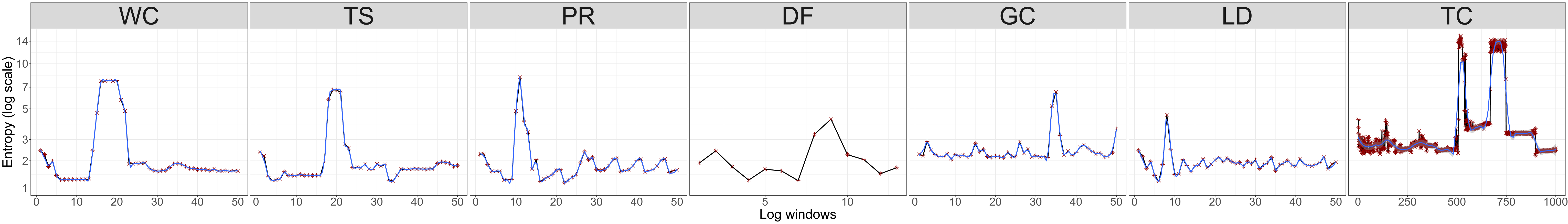}
\caption{Entropy values for log windows for different applications with Spark's compute node failure.}
\label{compute_node_failure}
\vspace*{-4mm}
\end{figure*}
  
\subsection{Compute Node Failure}\label{compute_failure}  
Figure~\ref{compute_node_failure} shows the entropy values over time for different benchmarks. 
The x-axis shows sequential log windows of 4 KBs in size, as suggested by prior work for log analysis time window (\cite{gholamian2021naturalness}), and the y-axis shows the entropies. 
As the size of the generated logs varies for each benchmark, we show the timeline of entropy changes for each benchmark from the start to the end of its execution. 
The spikes in the entropy values are the manifestation of failures in the Spark's logs. 
Once a failure happens, the system first detects the failure and then plans a set of \textit{recuperating actions} to recover from the failure. 
For example, for a distributed system such as Spark, the task manager resubmits the tasks previously assigned to a failed node to other nodes in the distributed cluster. 
This results in several log lines in the log files, which we call \textit{failure manifestation log region}, that have higher than normal entropy values. 
As the failure happens, the benchmarks also show noticeable prolonged computation time and additional logs compared to the baseline scenario (\textit{i.e.}, with no failure in Figure~\ref{ct_vs_so}).

\textbf{Detailed analysis of TransitiveClosure.} We observed that failures can result in different manifestations in the execution logs, and the manifestation can be relatively benchmark dependent. 
To provide further insight, we review the interesting scenario of a Spark compute node (CN) failure for the TransitiveClosure benchmark, which goes through the following four failure stages: 
\begin{enumerate*}[label=(\textbf{S{{\arabic*}}})]
\item \textbf{Failure detection}. Upon a CN failure, the Spark's Master (SM) observes this as \textit{``a CN has been disassociated''}, and subsequently, SM observes that the tasks associated with that CN are also lost. 
This results in a set of log records with high IG, and the first region of spikes in entropy values for \textbf{\textit{TC}} in Figure~\ref{compute_node_failure} right after \textit{x=500}.
\item \textbf{Interleaving logs}. In addition, due to the interleaving of logs in the distributed system, other components in the system still continue to generate normal logs. 
This is manifested in the entropy drop after the initial spike.  
\item \textbf{Recovery attempts.} Happens when SM makes several unsuccessful attempts to \textit{recover} from the failed state and reconnect to the lost CN. 
This is manifested as the second high entropy region in Figure~\ref{compute_node_failure} that ends just before \textit{x=750}. 
\item \textbf{Cleaning and back to normal}. 
After \textbf{S3}, SM gives up attempting to reconnect to the failed CN and continues the execution by reassigning the failed tasks to the remaining CNs. 
In the meanwhile, it clears the data structure allocated to the failed CN. 
\end{enumerate*} 
It should be noted that the outlined stages can manifest differently depending on the applications. 
For example, at \textit{info} level, TS generates far fewer log lines (0.57 MBs) than TC (313 MBs). 
As such, TS observes less interleaving of logs compared to TC, and \textbf{S1} and \textbf{S3} manifest as one spike region in the logs.

\begin{tcolorbox}[breakable, enhanced]
\small \textbf{Finding 4.} \textit{A compute node failure with manifestation in log files would result in higher entropy values than normal entropies, and different runs show extended computation time and additional logs related to the failure. 
CT of CPU-intensive applications suffers more from compute node failure than I/O intensive benchmarks. } 

\textbf{Implications.} \textit{Sudden changes in the entropy values of log records can signify a system failure.}
\end{tcolorbox}

\begin{figure}[h]
\centering
\includegraphics[width=.8\linewidth]{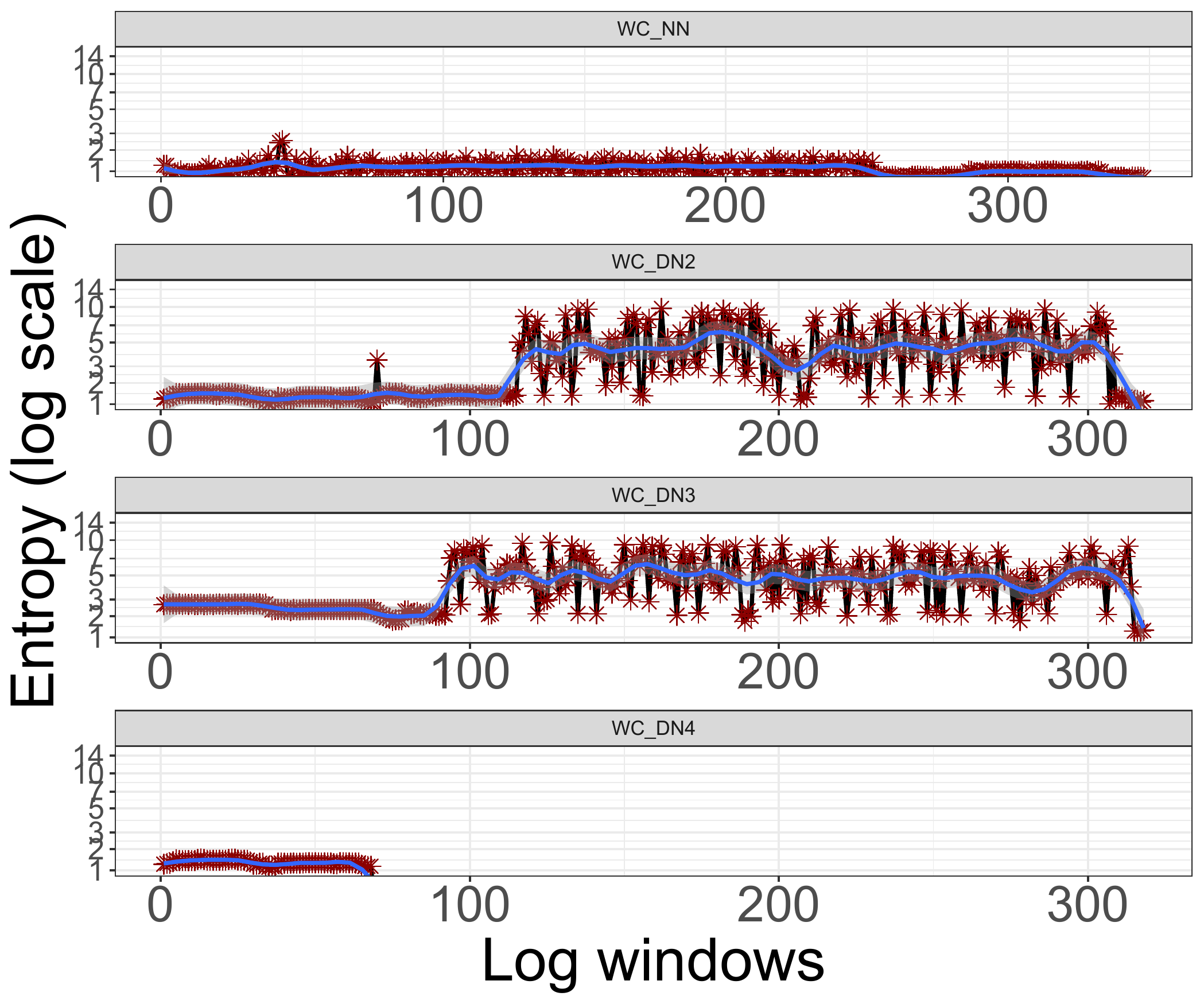}
\caption{Entropy values for log windows for WordCount with HDFS's data node failure.}
\label{entropy_dn_failure}
\vspace*{-6mm}
\end{figure}

\begin{figure}[h]
\centering
\includegraphics[width=.8\linewidth]{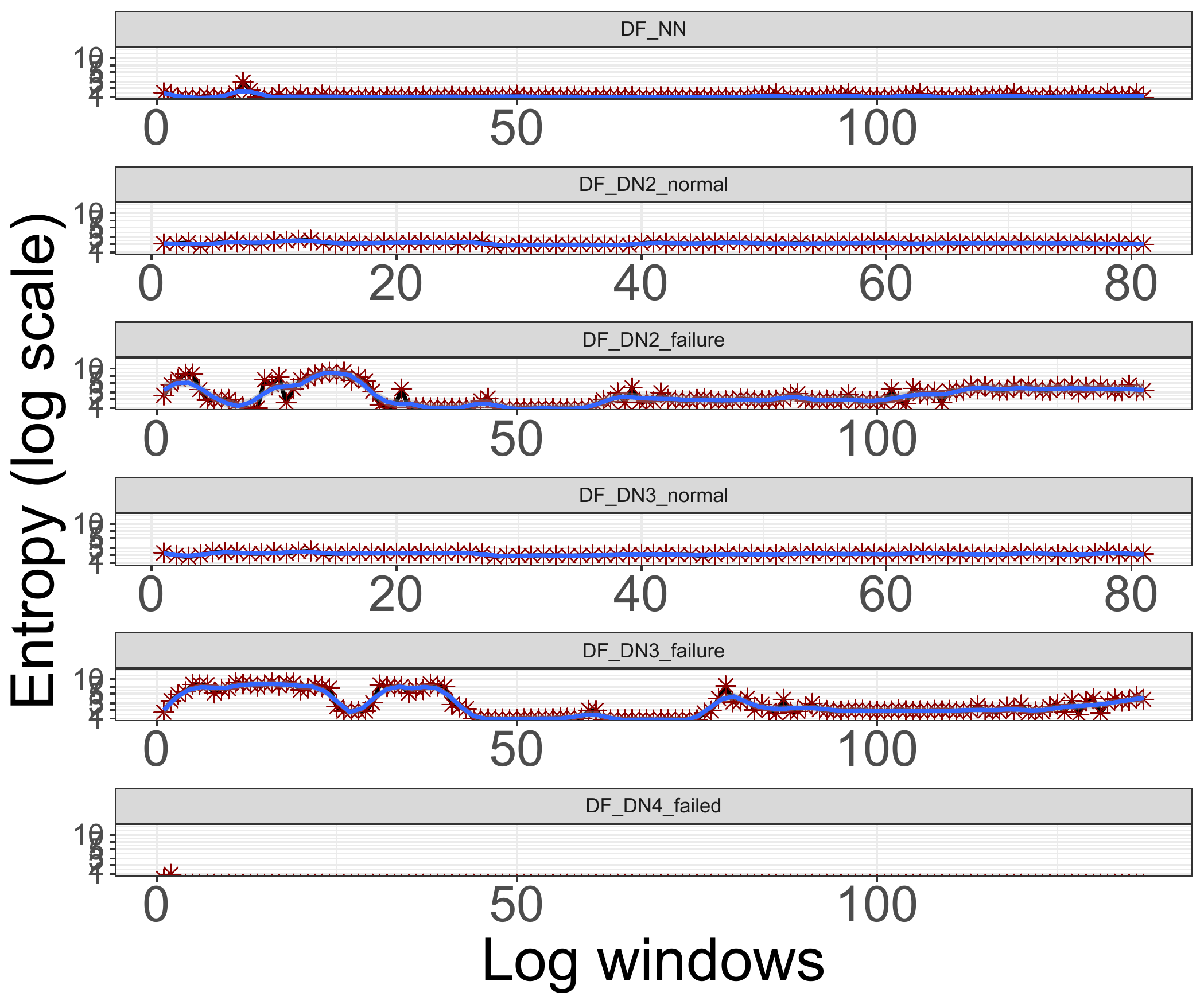}
\caption{Entropy values for log windows for DFSIO with HDFS's data node failure.}
\label{entropy_dn_failure_df}
\end{figure} 

\subsection{Storage Failure}\label{storage_failure}
To gain insight into storage failures, we investigate the entropy changes of HDFS logs. 
All nodes within the HDFS file system (\textit{i.e.}, name node and data nodes) generate logs. 
We perform the experimentation for all the benchmarks and review the logs from all the nodes and measure the entropy changes as the failure happens. 
Due to the limited space, we focus on the entropy value changes of WordCount and DFSIO as they are I/O intensive benchmarks, and they make the most use of HDFS compared to other benchmarks. 
Figure~\ref{entropy_dn_failure} shows the entropy values over time for the name node (WC\_NN), and three data nodes (WC\_DNx) as the failure happens. 
When DN4 fails at log window 69, we observe a delayed manifestation of entropy changes in other data nodes (DN2 and DN3) which starts at \textit{x=100}. 
We observe that DN2 and DN3 directly contact DN4 (which is not available) to retrieve some blocks of data, and hence this results in failure log messages and hence higher entropies.
By default, DNs are configured to send heartbeat signals to NN every 3 seconds. 
However, in case of a DN failure, NN marks an unresponsive DN dead after 10mm:30ss\footnote{\url{https://issues.apache.org/jira/browse/HDFS-3703}}, which at that time manifests in high entropy values in NN logs. 
The log windows in Figure~\ref{entropy_dn_failure} show the entire execution span for WC, which on average finishes within four minutes, and hence, we do not observe entropy value changes in NN in the plotted timeline. 
Similarly, Figure~\ref{entropy_dn_failure_df} represents the entropies for DFSIO, another I/O intensive task in our benchmark set when a data node (DN4) fails. 
We have also shown the entropy values for DN2 and DN3 during a normal run for comparison. The peaks show failure log messages with some normal interleaving logs. Failure for DN4 happens very close to the start of the x-axis and thus the initial peaks for DN2 and DN3.

\begin{tcolorbox}[breakable, enhanced]
\small \textbf{Finding 5.} \textit{We observe noticeable entropy changes in the HDFS logs of I/O intensive benchmarks as the storage failure occurs. 
CPU-intensive benchmarks that have minimal interaction with HDFS do not generate enough HDFS logs for a meaningful log analysis.} 

\small \textbf{Finding 6.} \textit{I/O intensive applications that read and write large volumes of data to the distributed storage will be negatively affected the most as a result of a storage node failure, whereas CPU-intensive applications (\textit{i.e.}, with minimal R/W to the storage) will be less impacted. Thus, CT of I/O tasks becomes prolonged due to the storage failure and the application's log size is also partially increased as it captures extra failure log records.} 

\textbf{Implications.} \textit{As failures are manifested as higher information gain, entropy-based anomaly detection approaches can be applied for online log analysis to isolate the higher entropy regions and further investigate the failures.}
\end{tcolorbox}

\subsection{Communication Interference Modeling}\label{com_interference}
\begin{figure}[h]
\vspace*{-5mm}
\centering
\includegraphics[width=.7\linewidth]{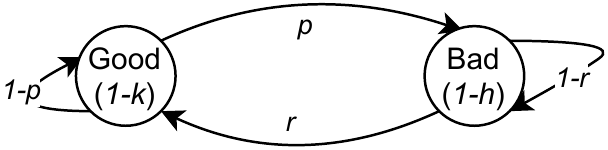}
\caption{Gilbert-Elliot communication interference model.}
\label{glmodel}
\vspace*{-2mm}
\end{figure}

If the network connection between distributed nodes permanently disconnects, the observable failure outcome would be similar to the permanent failure of the compute/storage node, as that node becomes unreachable from the cluster manager. 
In contrast to permanent failures in Sections~\ref{compute_failure}-\ref{storage_failure}, we here investigate intermittent network interference, \textit{e.g.}, packet loss, to gain insight into non-permanent failures which are manifested as \textbf{\textit{performance degradation}}. 
To emulate a realistic network traffic model, we implement Gilbert-Elliot capacity modeling approach~\cite{mushkin1989capacity,hasslinger2008gilbert}, which is comprised of \textit{Good} and \textit{Bad} states (Figure~\ref{glmodel}). 
This model offers a more realistic emulation for network impairments, rather than simple packet loss. 

An example usage configuration for Gilbert-Elliot scheme would be as follows: \texttt{``tc qdisc add dev dev\_name root netem loss gemodel 2\% 15\% 30\% 1\%.''} In this example, the error rate in Good \textit{(1−k)} and Bad \textit{(1−h)} states are \textit{1\%} and \textit{30\%}, respectively, and the probability of transitioning to Good \textit{(r)} and Bad \textit{(p)} states are \textit{2\%} and \textit{15\%}, respectively. 
In the following experiments, we vary the error rate in Bad state, \textit{i.e.,} \textit{(1−h)}, in the range of  (0\%, 45\%) and measure the computation time. 
In addition, we also evaluate the \textit{`combined failure'} (Case \textbf{4} in Section~\ref{rq3}) by disconnecting one of the machines in the cluster that hosts both compute and data nodes. 

\textbf{Heterogeneous cluster.} For the purpose of experimentation, we also define a new configuration that has three slave/data nodes but one node is smaller as it is using half of the cores and memory (\textit{i.e.}, 12 cores and 16 GBs of memory instead of 24 cores and 32 GBs of memory). 
This is in contrast to the homogeneous cluster (Figure~\ref{cluster_setup}) that all the three slave/data nodes are the same size (24 cores and 32 GBs of memory). 
The rationale to include the heterogeneous configuration is to compare it with the performance degradation scenario that appears as a result of network interference.

\begin{figure}
\begin{subfigure}{0.24\textwidth}
   \includegraphics[width=\linewidth]{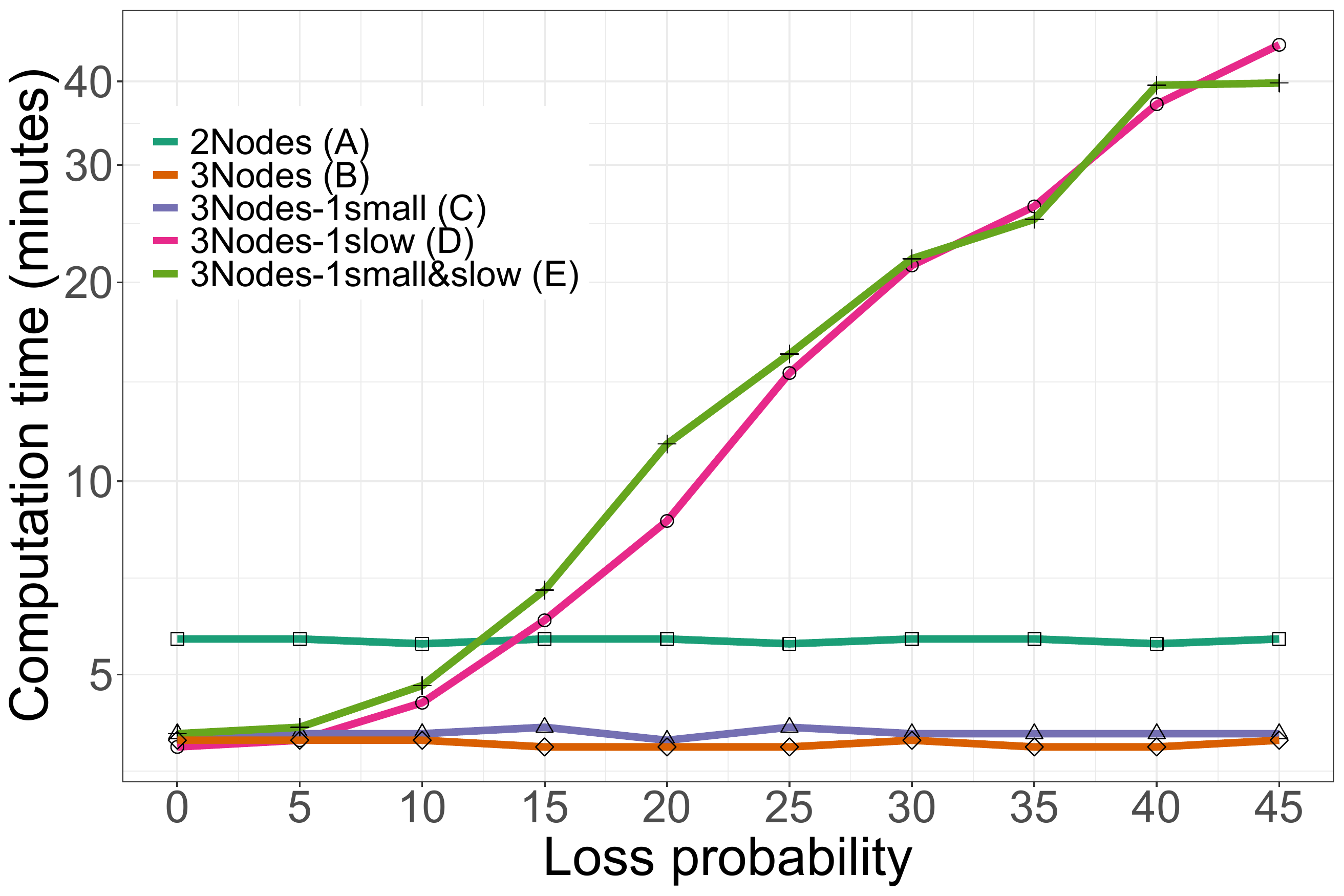}\vspace*{-2mm}
   \caption{WordCount} \label{entropy_dn_failure_wc}
\end{subfigure}
\begin{subfigure}{0.24\textwidth}
   \includegraphics[width=\linewidth]{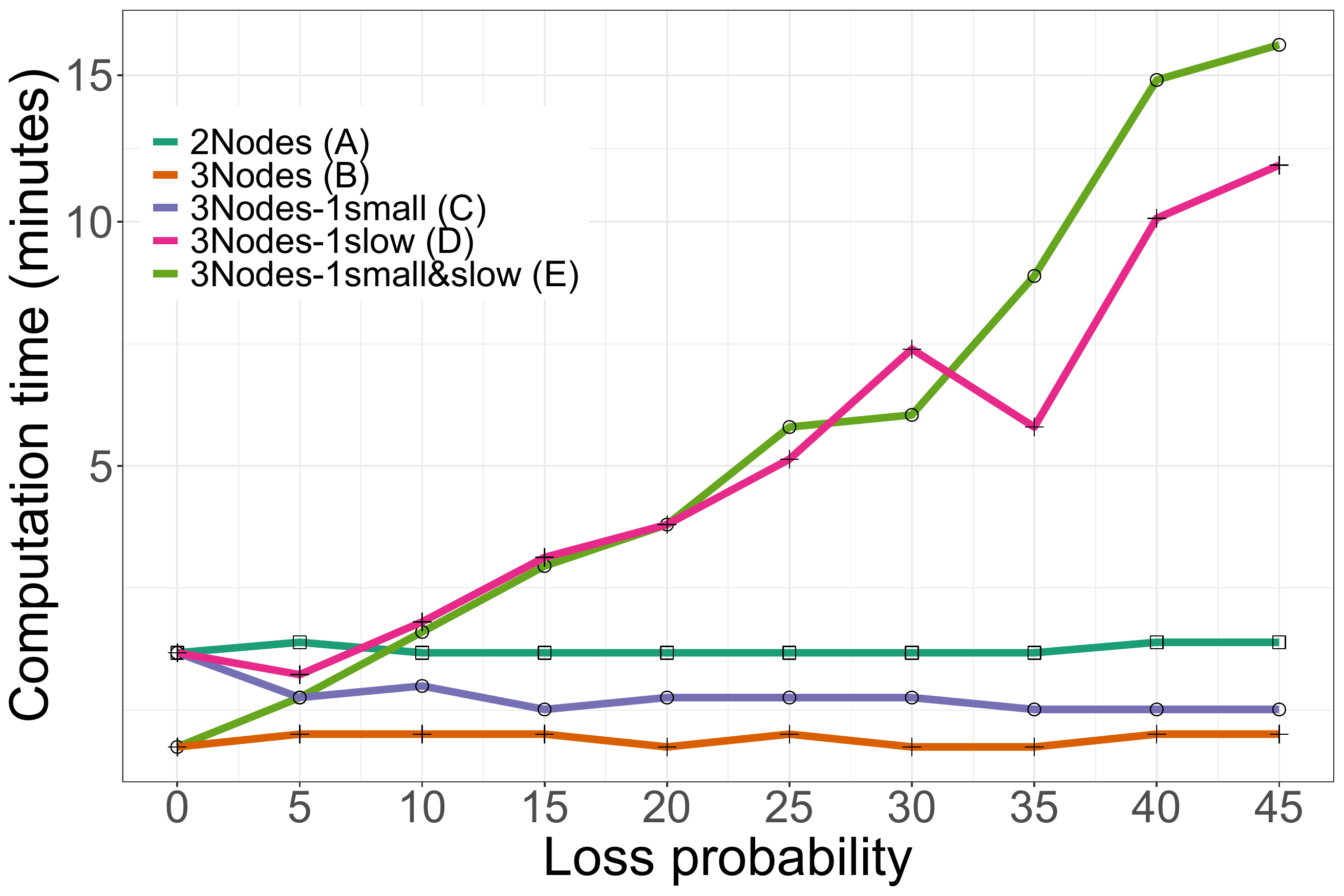}\vspace*{-2mm}
   \caption{TransitiveClosure} \label{entropy_dn_failure_tc}
\end{subfigure}
\caption{Execution time for WC and TC during the communication interference and combined failure.}
\label{hetergen_wc_tc}
\vspace*{-3mm}
\end{figure}

\textbf{Results.} Figures~\ref{entropy_dn_failure_wc} and~\ref{entropy_dn_failure_tc} show the evaluation for \textbf{communication interference} and \textbf{combined failure} for WordCount and TransitiveClosure as examples of I/O intensive, and iterative and CPU-intensive benchmarks, respectively. We refer to the graphs by their labels in the figures, \textit{i.e.}, (A)-(E).
Graph (A) shows the computation time as a result of a combined failure for a cluster with 2 nodes, \textit{i.e.}, two compute nodes and two storage nodes after a machine that hosts both compute and storage nodes fails. 
Graph (B) shows a \textit{homogeneous} cluster with three nodes, and Graph (C) shows a \textit{heterogeneous} cluster in which the third node is smaller (`3Nodes-1small'). 
No network interference is applied to (A), (B), and (C). 
Graphs (D) and (E) are equivalent to (B) and (C), respectively, but with added network interference. 
In Figure~\ref{hetergen_wc_tc}, as the communication interference increases, the CT time for (D) and (E) increases, and for values higher than 15\% for WC and 10\% for TC, the computation time of a cluster with interference surpasses a cluster with combined failure, Graph (A), \textit{i.e.}, two nodes in the cluster. 
We also observe that I/O intensive benchmarks that require to transfer a large amount of data among the nodes in the cluster suffer more than CPU-intensive benchmarks that use the network to a lesser degree.

\begin{tcolorbox}[breakable, enhanced]
\small \textbf{Finding 7.} \textit{As the communication interference increases, the computation time increases, and communication interference is manifested as a performance degradation and not a complete failure.} 

\small \textbf{Finding 8.} \textit{When the interference increases beyond a certain threshold, the negative impact of the performance degradation surpasses the impact of a complete failure because, for each stage of the computation, the faster nodes are awaiting the completion of the slow node.\looseness=-1} 

\textbf{Implications.} \textit{Distributed scheduling algorithms that can detect slow nodes in the system and remove them from the computation can benefit the entire system's performance.}
\end{tcolorbox}

\begin{figure}
\centering
\vspace*{-2mm}
\includegraphics[width=.99\linewidth]{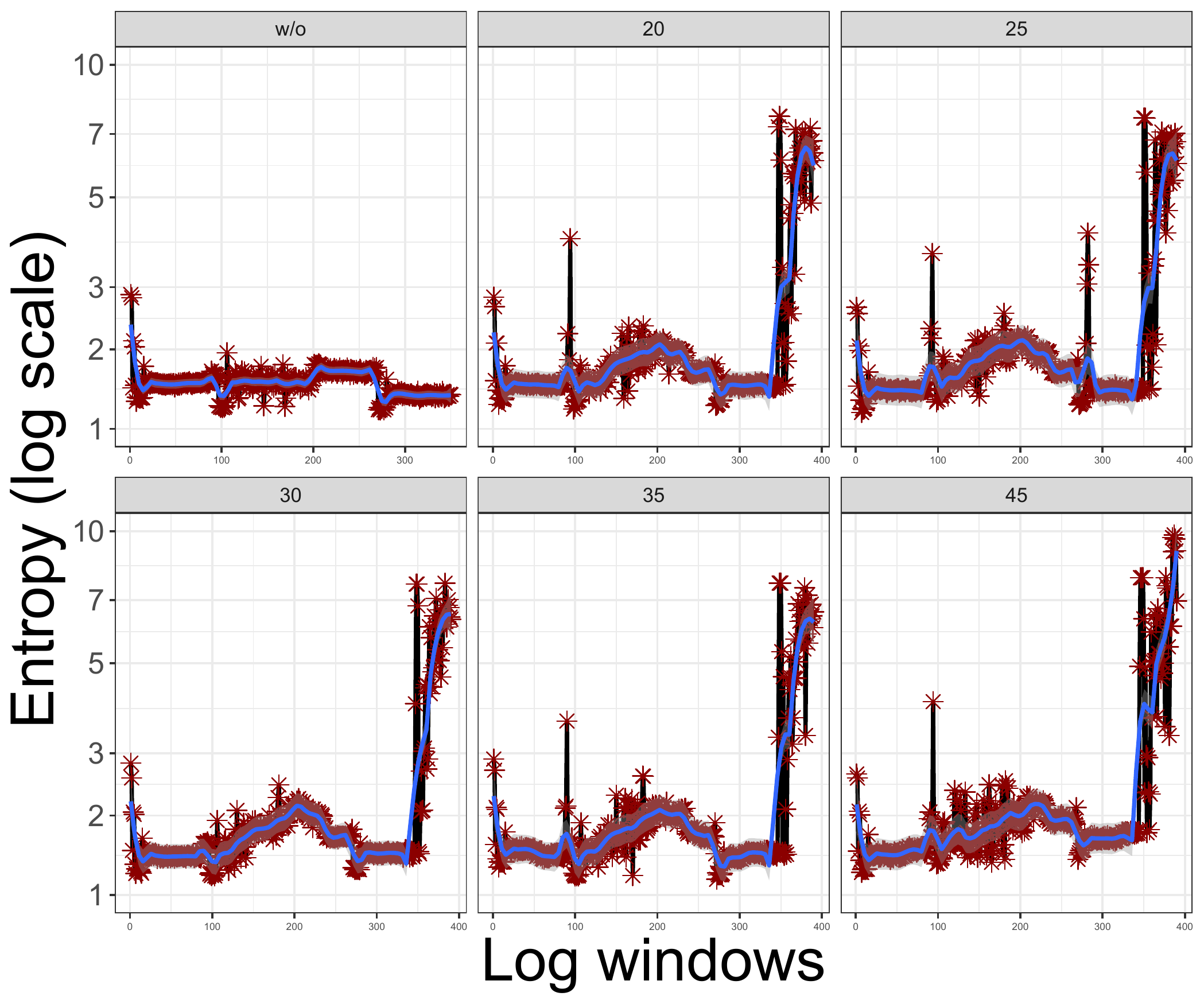}
\caption{Entropy values for log windows for WordCount for different values of drop rate \textit{(1-h)}.}
\label{entropy_tc_wc}
\vspace*{-5mm}
\end{figure}

\textbf{Entropy values.} Figure~\ref{entropy_tc_wc} shows the effect of drop rate in Bad state \textit{(1-h)} for WordCount logs. 
The top-left graph (`w/o') shows the entropy of Spark's logs with zero communication interference, and we gradually increase the drop rate from 5\% to 45\% (in steps of 5\%) and plot the entropy values from the start to the end of the execution for selected percentages, \textit{i.e.}, (20, 25, 30, 35, and 45)\%. 
Our observation is that due to the non-deterministic nature of network interference, performance degradation is indirectly manifested in the logs and their corresponding entropy values, in contrast to having clear regions with high entropies (Figures~\ref{compute_node_failure}-\ref{entropy_dn_failure}).
We also observe that entropy values start to climb as the interference percentage increases. 
In addition, higher entropy values are manifested with a delay towards the end of the execution as the system experiences timeouts and aims to reestablish the connection with the unstable node or resubmit the failing tasks to other nodes in the distributed system. 
Therefore, we hypothesize that a combination of execution log records and system metrics, such as average task completion time for speculative execution (Section~\ref{discussion}), are required to identify performance degradation cases~\cite{ibidunmoye2015performance}.

\begin{tcolorbox}[breakable, enhanced]
\small \textbf{Finding 9.} \textit{As the communication interference increases, the entropy values gradually increase with a delay.} 

\textbf{Implications.} \textit{Failures that manifest as a gradual system slowdown and performance degradation are harder to detect than complete failures solely with logs. 
As such, other system metrics, \textit{e.g.}, average task completion time, can be applied in conjugation with logs.}
\end{tcolorbox}

\subsection{Discussion}\label{discussion}
\begin{figure}
\begin{subfigure}{0.24\textwidth}
   \includegraphics[width=\linewidth]{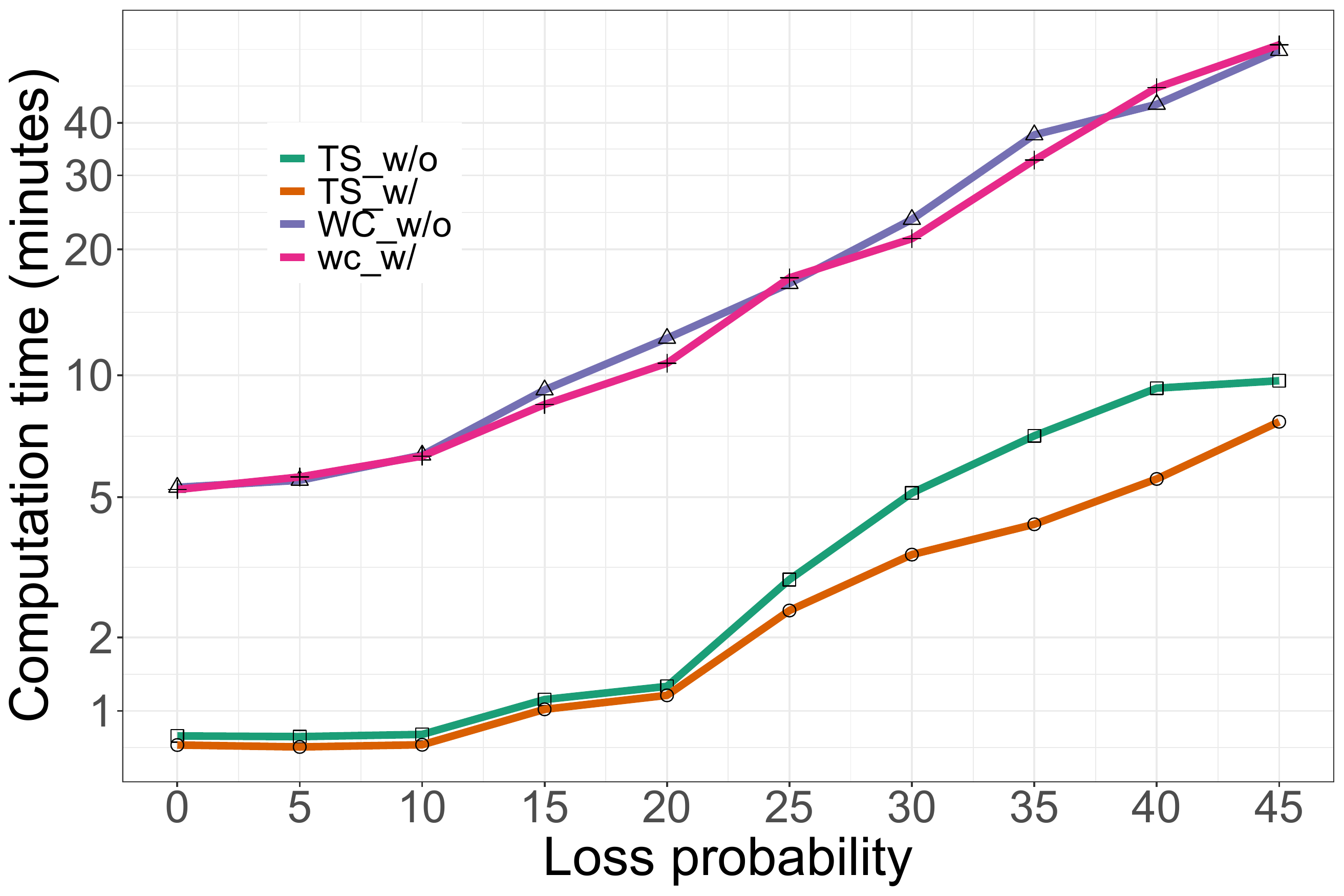}\vspace*{-2mm}
   \caption{WC and TS} \label{sponly_wc_ts}
\end{subfigure}
\begin{subfigure}{0.24\textwidth}
   \includegraphics[width=\linewidth]{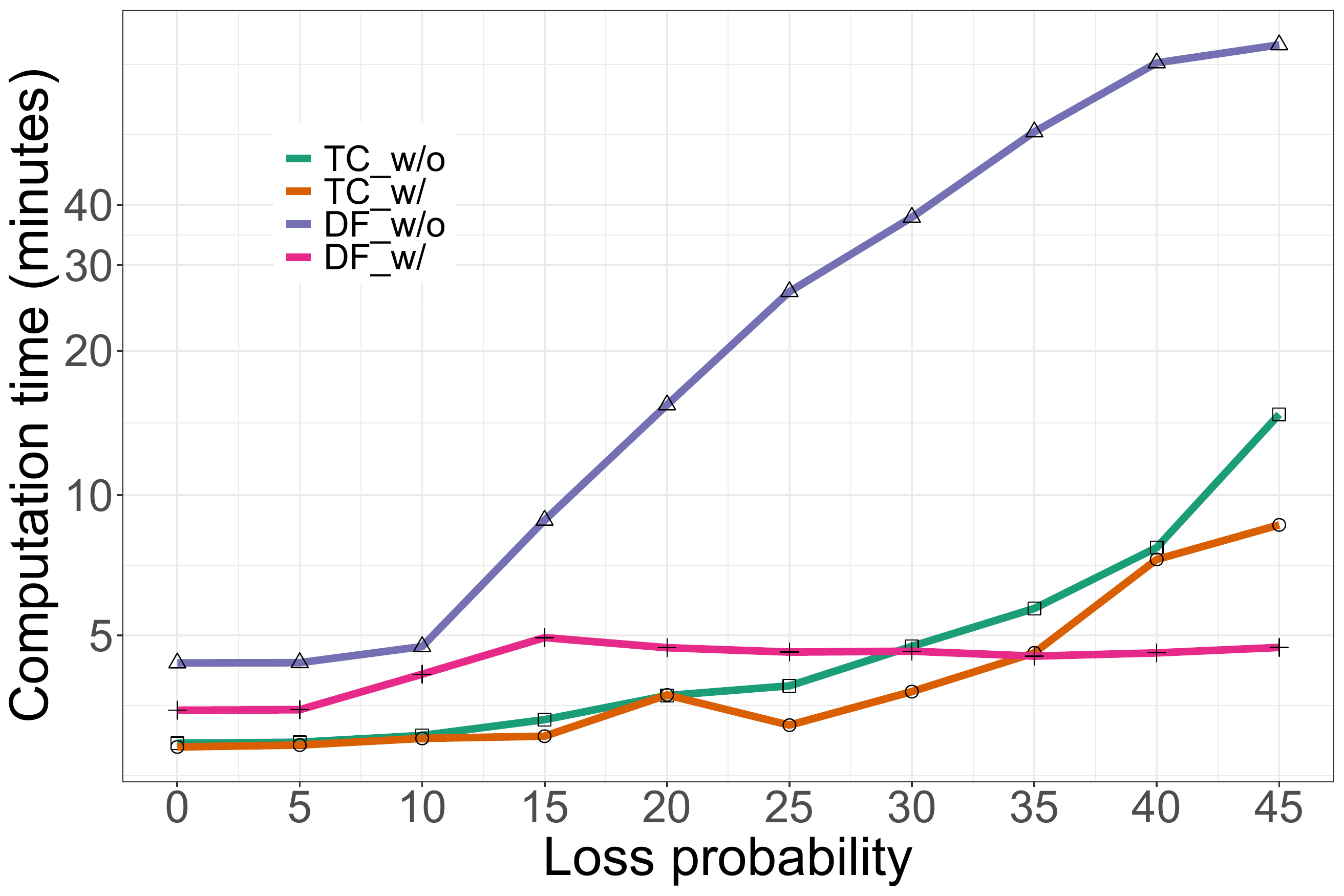}\vspace*{-2mm}
   \caption{TC and DF} \label{sponly_tc_df}
\end{subfigure}
\caption{Speculative execution for different benchmarks.}
\label{hetergen_spoly}
\vspace*{-6mm}
\end{figure}

\textbf{Speculative execution.} Findings 7-9 imply that since communication interference is manifested as intermittent failures, as opposed to a complete compute or data node failure, they are harder to detect and diagnose from the log files.  
Therefore, we suggested the usage of distributed scheduling algorithms that can detect the slow nodes and utilization of system metrics (\textit{e.g.}, average task's computation time) in conjunction with logs for more effective problem diagnosis. 
Spark provides a feature known as speculative execution (\texttt{spark.speculation})~\cite{urlsparkspeculative} that if enabled, allows resubmitting slow tasks to other nodes in parallel, and proceeds as soon as any of the task instances completes its execution. 
Figure~\ref{hetergen_spoly} shows the non-speculative (`w/o') and speculative (`w/') runs for WC, TS, TC, and DF. 
For WC, although the computation is moved to another node, because large amounts of data are being shuffled through the network interference to the new node, speculation would not show a noticeable improvement. 
For DF, we observe as the network interference increases, without speculative execution, the write pipeline\footnote{\url{https://stackoverflow.com/questions/37531946/what-is-hdfs-write-pipeline}} fails more often, and this results in a noticeable increase in the execution time of this benchmark. 
With speculative execution, the slow tasks are moved to other nodes, hence with a different write pipeline, which significantly helps with containing the network failures. 
In short, the main benefit from speculative execution in the distributed environment comes from moving tasks from the faulty node with network interference to other nodes, which helps to avoid task re-execution and data re-transfer due to the network uncertainty. 
Our observation is that speculative execution only marginally improves the execution time, and in general, after a certain level, even with speculative execution enabled, the distributed systems performance becomes slower than removing the node with lower performance completely from the cluster.\looseness=-1 

\textbf{Cluster heterogeneity.} 
We experimented with homogeneous and heterogeneous clusters in Section~\ref{com_interference}. 
Although heterogeneity by design, \textit{i.e.}, having a machine smaller than other machines in the cluster, is well understood~\cite{page2010multi}, heterogeneity that is the result of performance degradation and partial failure is left untreated. 
In our case, the rationale to include experiments with a heterogeneous cluster is to simulate a scenario that heterogeneity is introduced in the distributed platform because of performance degradation. 
In other words, although the original design is homogeneous, heterogeneity can still happen due to a variety of reasons, such as communication interference, which can negatively impact the entire system's performance.  
Also, one of the factors that limits Spark speculation performance is its assumption about operating in a homogeneous environment, which is not the case in a performance-impaired cluster. This would encourage further research to investigate possible scenarios and solutions for failure-induced heterogeneity.\looseness=-1

\textbf{Slow distributed file system.}
Although with the replication factor of three in HDFS, there is no single point of failure, partial failures can negatively affect the performance of the entire file system.
In our case, the slow network connection for one of the data nodes due to an induced drop rate negatively impacts the performance of the entire HDFS. 
A slow data node still continues to send heartbeats successfully, and the HDFS name node will keep redirecting clients to the slow DN, and therefore, degrade the performance of the entire cluster. 
Although HDFS provides few settings to detect and report slow data nodes, it does not provide a mechanism to automatically bypass the slow DNs, as they are still sending heartbeat signals to the name node. 
Thus, we foresee further research to investigate mechanisms, similar to Spark's speculative execution, to obtain data from other available data nodes in case a data node becomes intermittently unavailable or slow.    

\textbf{Implications on fault tolerance.} Our observation is that information gain is helpful in zooming in the failure regions (\textit{i.e.}, spikes in the entropy values), which means that these regions of logs with higher IG can be isolated and quickly reviewed by users and practitioners for failure diagnosis, which then results in faster system recovery from a failure. 
Additionally, we emphasize that although the spark cluster is fault-tolerant by design, its performance is impaired due to the failures. Thus, we envision that quicker detection and recovery of failed compute and storage nodes directly connects with and benefits the robustness and fault-tolerance of the system. 
A speedy return to the normal state will allow the system to tolerate additional failures.\looseness=-1

\section{Case study}\label{case_study}
\begin{figure}[t]
\centering
\includegraphics[width=.99\linewidth]{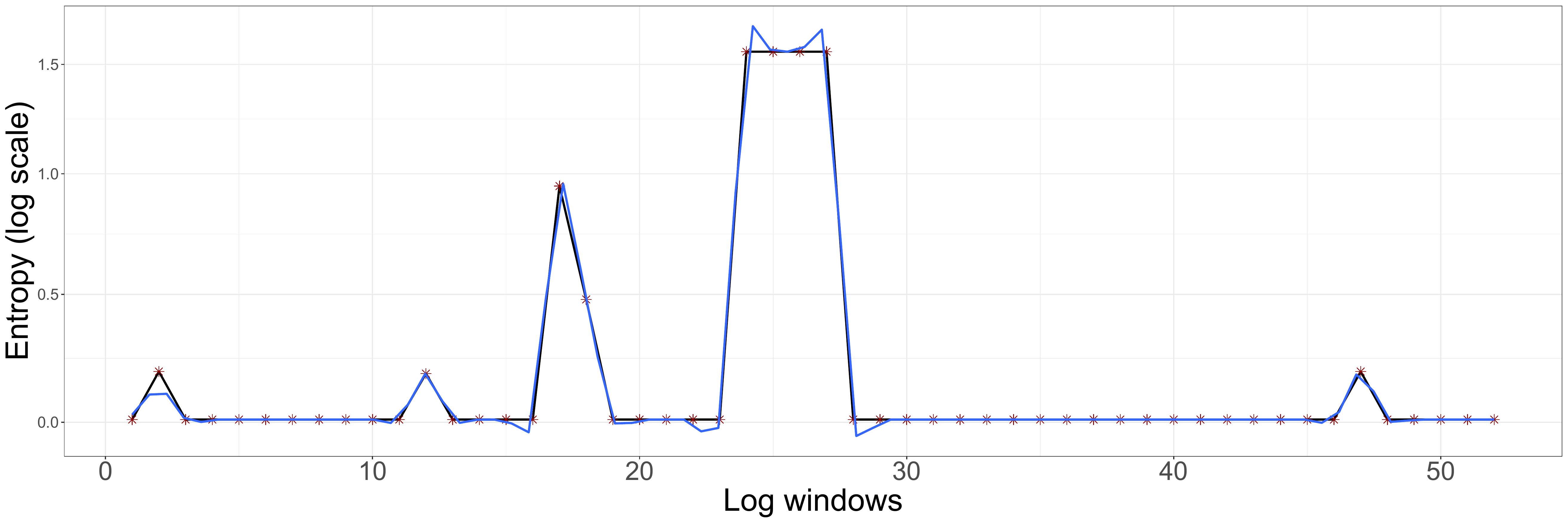}
\caption{OpenStack entropy values for log sequences with four anomalous VM log records.}
\label{openstack_anomaly}
\vspace*{-5mm}
\end{figure}

In this section, we use logs from real abnormal scenarios for labeled OpenStack logs~\cite{du2017deeplog}. 
We first parse the logs to extract their templates and then group the logs based on their \textit{instance\_id}. 
\textit{instance\_id} serves as an identifier for a particular task execution sequence. 
Figure~\ref{openstack_anomaly} shows the timeline of 52 log windows with 4 injected abnormal OpenStack VM sequences in the center for $x \in [24, 27]$. 
There is also one false positive high entropy value at $x=17$. By leveraging the \textit{Hampel} filter and outlier detection approach proposed in prior work~\cite{gholamian2021naturalness}, we can reach \textit{F-Measure} ($\frac{2\times precision \times recall}{precision+recall}\vspace{1mm}$) and \textit{Balanced Accuracy} $ (\frac{TP}{2(TP+FN)} + \frac{TN}{2(TN+FP)})\vspace{1mm}$ of 0.89 and 0.98, respectively. This signifies the effectiveness of the information gain approach that is achieved by measuring entropy values of log sequences in detecting OpenStack's abnormal scenarios.

\textbf{Computation time.} In our analysis, the training of the natural language model happens only once on normal logs. 
Then, testing the log records while they are generated is rather fast. 
We performed a quick measurement and received on average \textit{2.4} milliseconds execution time (as a single thread executed on 2.40GHz Intel Xeon E5-2620) for a \textit{4KB} log window. 
Relatively, the execution time for testing logs is faster than the rate of log generation, thus, the log records can be tested in real-time and observed for information gain and potential anomalies.

\section{Related work}\label{rwork}
\textbf{Logging cost and gain.} 
Prior work on assessing logging cost does not quantitatively evaluate the system performance overhead contributed to logging levels. 
Ding et al.~\cite{ding2015log2} performed a survey of logging practices among Microsoft developers and listed their findings on overheads associated with logging reported by the developers.  
Mizouchi et al.~\cite{mizouchi2019padla} presented PADLA, an online method to dynamically adjust the logging level, and, consequently, limit the logging cost. 
Miranskyy et al.~\cite{miranskyy2016operational} reviewed challenges for log analysis of big-data systems, among them \textit{limited storage} and \textit{unscalable log analysis}. 
Goshal \textit{et al.}~\cite{ghoshal2013provenance} discussed the provenance, \textit{i.e.}, the origin, of logs and how it correlates with log levels and types of applications in large-scale systems. 
Different from the prior research, our study aims to quantitatively analyze the costs and benefits associated with the level of logging and the information gain.\looseness=-1

\textbf{Spark's performance evaluation.}
Spark~\cite{zaharia2012resilient} has been introduced for massively parallel data analytics, which improves on its predecessor, Hadoop MapReduce~\cite{dean2008mapreduce}, by utilizing in-memory storage of intermediate results for iterative applications, and bringing more flexibility in performance and the programming model~\cite{shi2015clash,petridis2016spark}. 
Mavridis and Karatza~\cite{mavridis2017performance} evaluated various log file analyses with the cloud computational frameworks, Apache Hadoop and Apache Spark, and, experimentally, Spark achieved the best performance. 
Lu \textit{et al.}~\cite{lu2017log} performed log-based anomaly detection for Spark. 
Our study is different as we leverage Spark's logs to evaluate the cost and IG associated with log levels.

\textbf{Information gain and NLP.} 
In this research, we propose a new perspective along with validated metrics to evaluate the impact of different verbosity levels on cost and information gain from log statements with natural language processing (NLP) approaches. 
Compared with related research~\cite{anu2019approach,li2017log,gholamian2020logging}, our approach is orthogonal to such efforts that aim to suggest the proper logging statement or its VL by extracting features from the source code. 
These works rely on the existing logging statements to suggest logs for newly composed instances of code. 
However, we aim to bring attention to the trade-offs between logging cost and the information gain, and that failures are manifested as higher information gain in logs.\looseness=-1 

\section{Conclusions and Future Work}\label{conclusion}
The goal of our work is to provide a quantitative assessment of logging cost in different verbosity levels and how that translates to information gain in distributed systems.
Therefore, we evaluate the impact of log verbosity levels on performance and storage overhead, and the information gain from logs for various Spark Benchmarks.
We also experiment with synthesizing various categories of distributed failures for compute and data nodes, and network interference, and measure the effect of failures in execution time, the volume of the generated logs, and the information gain. Lastly, we provide a case study of the application of our approach on OpenStack real failure logs. 
Our findings are helpful for developers and practitioners to better evaluate the costs and benefits of logging when choosing different verbosity levels and how failures can be tracked down with IG approaches.  
As future work, we will look into evaluating logs of other distributed software systems and investigate how IG can be translated to more effective troubleshooting by leveraging the execution logs and system metrics.

\bibliographystyle{abbrv}
\bibliography{logging_cost_ieee_SRDS2021}

\end{document}